\documentclass[aps,prd,showpacs,nofootinbib,amsmath,superscriptaddress]{revtex4}\usepackage{bm}
\usepackage{graphicx}
\newcommand{\de}{{\rm d}}

\begin{document}

\title{Collins and Sivers effects in $p^{\uparrow}\,p\to {\rm jet}\,\pi\,X$:
Universality and process dependence}

\author{Umberto D'Alesio}
 \email{umberto.dalesio@ca.infn.it}
 \affiliation{Dipartimento di Fisica, Universit\`a di Cagliari, Cittadella Universitaria,
 I-09042 Monserrato (CA), Italy}
 \affiliation{Istituto Nazionale di Fisica Nucleare, Sezione di Cagliari, C.P. 170,
 I-09042 Monserrato (CA), Italy}

\author{Francesco Murgia}
 \email{francesco.murgia@ca.infn.it}
 \affiliation{Istituto Nazionale di Fisica Nucleare, Sezione di Cagliari, C.P. 170,
 I-09042 Monserrato (CA), Italy}

\author{Cristian Pisano\footnote{Present address:~Nikhef and Department of Physics and Astronomy,
VU University Amsterdam, De Boelelaan 1081, NL-1081 HV Amsterdam, The Netherlands}}
 \email{cristian.pisano@ca.infn.it}
  \affiliation{Dipartimento di Fisica, Universit\`a di Cagliari, Cittadella Universitaria,
 I-09042 Monserrato (CA), Italy}
 \affiliation{Istituto Nazionale di Fisica Nucleare, Sezione di Cagliari, C.P. 170,
 I-09042 Monserrato (CA), Italy}

\date{\today}

\begin{abstract}
In this paper we briefly review the transverse momentum dependent generalized parton model and its
application to the study of azimuthal asymmetries in the distribution of leading hadrons (mainly
pions) inside large transverse momentum jets inclusively produced in polarized proton-proton collisions.
We put particular emphasis on the phenomenological interest of these observables, in
combination with similar asymmetries measured in semi-inclusive deeply inelastic scattering,
Drell-Yan processes and $e^+e^-$ collisions, for the study of the universality properties
of the transverse momentum dependent parton distribution and fragmentation functions.
We present results for RHIC kinematics at center-of-mass energies $\sqrt{s}= 200$ and 500 GeV, for
central and mainly forward jet rapidities, in particular for the Sivers distribution and
the Collins fragmentation function, that are believed to be responsible of many of the
largest asymmetries measured in the last years.
We also briefly discuss the case of inclusive jet production and
recent phenomenological applications of other theoretical approaches,
like the colour gauge invariant generalized parton model and the collinear twist-three approach,
aiming at clarifying the issues of the universality and process dependence
of transverse momentum dependent functions.
\end{abstract}


\maketitle

\section{\label{intro} Introduction}
In the last years the study and knowledge of the full three-dimensional dynamical
nucleon structure in polarized high energy collisions have witnessed impressive progress
(see e.g.~Refs.~\cite{D'Alesio:2007jt,Barone:2010zz} for recent reviews).
Motivated by several experimental results on spin and azimuthal asymmetries,
a class of partonic, transverse momentum dependent, distribution and fragmentation
functions (nowadays largely known as TMDs for short) have been introduced and analyzed.
In high energy hadronic processes where two energy scales play a role (a large perturbative scale
and a small transverse momentum scale) the usual leading-twist QCD collinear factorization schemes,
making use of the corresponding collinear parton distribution (PDFs) and fragmentation (FFs) functions,
often fail to describe several puzzling experimental measurements on spin asymmetries.
When a small transverse scale is involved one needs to take care more accurately
of the intrinsic motion of constituent partons inside parent hadrons.
Typical examples are: the low transverse momentum distribution of dilepton pairs
in Drell-Yan (DY) processes and the corresponding asymmetries in the azimuthal distribution
of the observed pair~\cite{Tangerman:1994eh,Boer:1999mm,Anselmino:2002pd};
the low transverse momentum spectrum of hadrons produced in the current region
in semi-inclusive deeply inelastic scattering (SIDIS)~\cite{Mulders:1995dh,Boer:1997nt,Anselmino:2011ch};
the azimuthal asymmetries in the correlations of two leading hadrons (typically pions) observed in opposite
jets produced in $e^+e^-$ collisions~\cite{Boer:1997mf,Anselmino:2007fs}.
Despite the theoretical and experimental difficulties associated with the study of these
reactions, they offer a unique opportunity to learn about the hadron structure in
the transverse directions (with respect to the usual light-cone one).

{}From an historical perspective, the first sizable single spin asymmetries were observed
in single inclusive hadron production at large values of the Feynman variable,
$x_F=p_L/p_L^{\rm max}\simeq 2p_L/\sqrt{s}$, and moderately large transverse momentum
in polarized hadronic collisions.
However, the theoretical study of this process is made difficult by the fact that
there is no small transverse momentum scale. Intrinsic transverse momenta
of partons are integrated out in the observable. This complicates the treatment of
such (higher twist) asymmetry, since several possible effects are mixed up and it is not obvious
how to disentangle them. Moreover, a TMD factorization scheme similar to that developed for the reactions
discussed above (DY, SIDIS, $e^+e^-$ annihilations) has never been proved and,
at least for double inclusive jet and/or hadron production processes,
like the one studied here, there are clear indications
that factorization may be broken (for a recent discussion, see e.g.~Ref.~\cite{Rogers:2013zha}
and references therein).

Quite recently, it has been suggested to study azimuthal asymmetries in hadronic collisions
by looking at the azimuthal distribution of leading hadrons (pions or kaons) inside a
large transverse momentum jet inclusively produced in polarized
proton proton collisions~\cite{Yuan:2007nd,D'Alesio:2010am}.
Although also in this case a proof of TMD factorization is not available (if one takes
into account intrinsic motion in the initial colliding hadrons), the observables considered
are rather similar to those measured in SIDIS. In particular, leading-twist
asymmetries appear and different contributions (like the Sivers or Collins effects) can be
disentangled by taking appropriate moments of the azimuthal distributions, much in the
same way adopted in the SIDIS or DY cases.

The detailed analysis of this process can be of crucial relevance, when compared with
analogous studies in the DY and SIDIS cases, for the theoretical and phenomenological
understanding of the process dependence and the universality properties of the
Sivers distribution and the validation of the expected universality of the Collins
fragmentation function. Other TMDs can also be tested in the same way.

In this review we summarize recent results concerning the study of
the Sivers and Collins azimuthal asymmetries in the distribution of leading pions
inside a jet in $p^\uparrow p\to {\rm jet}\, \pi\, X$ processes.
After a short description of the TMD theoretical approach adopted, the so-called generalized parton
model (GPM)~\cite{D'Alesio:2004up,Anselmino:2005sh}, we present
a selection of interesting results involving the Sivers and Collins effects,
that are expected to be the dominant contributions to the single spin asymmetries considered here.

We will also discuss in some details an extension of the GPM~\cite{Gamberg:2010tj},
named colour gauge invariant (CGI) GPM,
including colour gauge factors in the approach,
and its application to the study of the process dependence of the
Sivers distribution~\cite{D'Alesio:2011mc}. This is indeed expected in perturbative QCD, due to the essential
role played by initial and final state interactions among active partons and parent
hadrons for the nonvanishing of these single-polarized observables.

Finally, we will shortly summarize other recent attempts to study the process dependence
of the Sivers distribution (and other TMD PDFs and FFs) in different processes and adopting different
theoretical approaches. Hopefully, the combined phenomenological
analysis of several reactions and observables will help in clarifying
essential theoretical issues crucial for a full understanding of these interesting
phenomena in the realm of QCD.

\section{\label{kinematics} Kinematics}
\begin{figure*}[t]
 \includegraphics[angle=0,width=0.8\textwidth]{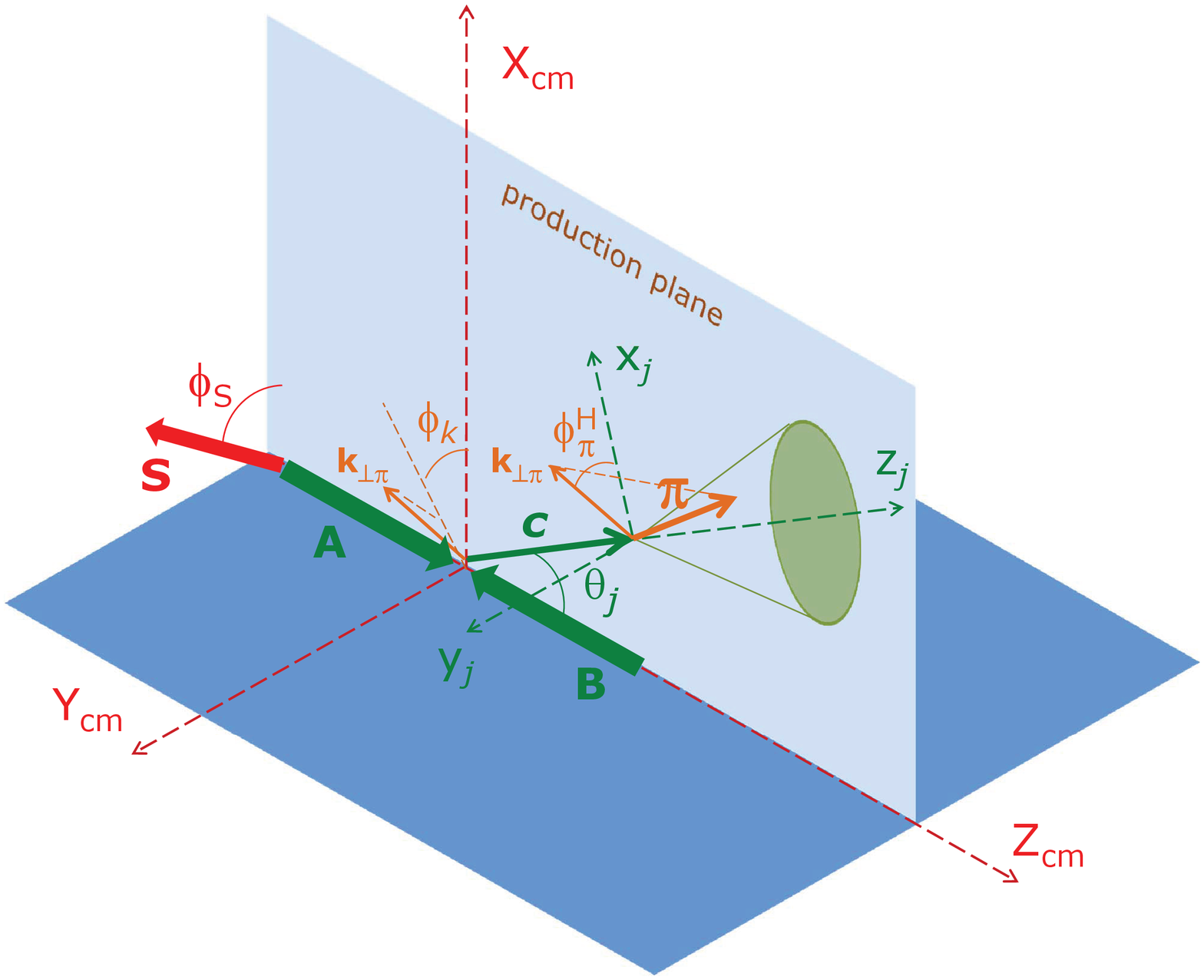}
 \caption{Kinematics for the
 process $A(p_A;S)+B(p_B)\to {\rm jet}(p_{\rm j})+\pi(p_\pi)+X$ in the
  center-of-mass frame of the two incoming hadrons, $A$ and $B$.
 \label{fig-kinem} }
\end{figure*}

We consider the process
\begin{equation}
A (p_A; S) \,+\, B (p_B)\, \to \, {\rm jet}(p_{\rm j})\,
+\pi(p_\pi)\, +\, X\,,
\end{equation}
where  $A$ and $B$ are two spin-1/2 hadrons carrying momenta $p_A$ and
$p_B$ respectively. One of the two hadrons, $A$,  is in a pure transverse
spin state described by the four-vector $S$ ($S^2=-1$ and $p_A\cdot S =0$), while $B$ is unpolarized.
We work mainly in the center-of-mass (c.m.) frame of $A$ and $B$,
where
$s = (p_A+p_B)^2$ is the total energy squared, and, as depicted in Fig.~\ref{fig-kinem},  $A$ moves along the  positive direction of the $\hat{\bm{Z}}_{\rm cm}$ axis. The production plane containing the colliding beams and the observed jet
is taken as the $(XZ)_{\rm cm}$ plane,
with  $(\bm{p}_{\rm j})_{X_{\rm cm}}>0$. In this frame the four-momenta of the
particles and the spin vector $S$ are given by
\begin{eqnarray}
p_A &=& \frac{\sqrt{s}}{2}(1,0,0,1)\,,  \qquad S \,= \,S_T \,= \,(0,\cos\phi_{S},\sin\phi_{S},0)\,,\nonumber\\
p_B &=& \frac{\sqrt{s}}{2}(1,0,0,-1)\, ,\nonumber\\
p_{\rm j} & = & (E_{\rm j}, p_{{\rm j}T},0,p_{{\rm j}L})\, =\,
E_{\rm j}(1,\sin\theta_{\rm j},0,\cos\theta_{\rm j}) \, = \,
p_{{\rm j}T}(\cosh \eta_{\rm j},1,0,\sinh \eta_{\rm j})\,, \nonumber\\
p_{\pi} &=&
E_{\pi}(1,\sin\theta_{\pi}\cos\phi_\pi,\sin\theta_{\pi}\sin\phi_\pi,\cos\theta_{\pi})\,,
\label{4mom-cm}
\end{eqnarray}
where all masses have been neglected and  $\eta_{\rm j}$ denotes the jet
(pseudo)rapidity, $\eta_{\rm j} = -\log[\tan(\theta_{\rm j}/2)]$.

At leading order in perturbative QCD, the reaction proceeds
via the partonic hard scattering subprocesses $ab\to cd$, where the outgoing
parton $c$   fragments into the observed hadronic jet.
For the partonic
momenta in the hadronic c.m.\ frame, one has
\begin{eqnarray}
p_a &=& \left(x_a\frac{\sqrt{s}}{2}+\frac{{k}_{\perp a}^2}{2x_a\sqrt{s}},
k_{\perp a}\cos\phi_a,k_{\perp a}\sin\phi_a,
x_a\frac{\sqrt{s}}{2}-\frac{{k}_{\perp a}^2}{2x_a\sqrt{s}}\right)\,,\nonumber\\
p_b &=& \left(x_b\frac{\sqrt{s}}{2}+\frac{{k}_{\perp b}^2}{2x_b\sqrt{s}},
k_{\perp b}\cos\phi_b,k_{\perp b}\sin\phi_b,
-x_b\frac{\sqrt{s}}{2}+\frac{{k}_{\perp b}^2}{2x_b\sqrt{s}}\right)\,,\nonumber\\p_c & \equiv & p_{\rm j}\,,
\label{4mom-cm-par}
\end{eqnarray}
where  $k_{\perp a,b} = \vert \bm{k}_{\perp a,b} \vert$. Here we have
introduced the variables $x_{a,b}$ and $\bm{k}_{\perp a,b}$,
which are, respectively, the light-cone
momentum fractions and the intrinsic transverse momenta of the incoming partons $a$ and $b$. From Eqs.~(\ref{4mom-cm}) and (\ref{4mom-cm-par}) one can calculate the  partonic Mandelstam variables:
\begin{eqnarray}
\hat s  =  (p_a+p_b)^2 & = &  x_a x_b s \left[1 - 2 \left(\frac{k_{\perp a}
    k_{\perp b}}{x_ax_b s}\right) \cos(\phi_a-\phi_b) +
\left(\frac{k_{\perp a} k_{\perp b}}{x_a x_b s}\right)^2
\right] \,,\\
\hat t  =  (p_a-p_c)^2 & = &  - x_a E_{\rm j}\sqrt{s}\,\left[ 1-\cos\theta_{\rm j} -
2 \left(\frac{k_{\perp a}}{x_a\sqrt{s}}\right) \sin\theta_{\rm j}\cos\phi_a +
\left(\frac{ k_{\perp a}}{x_a\sqrt{s}}\right)^2 (1+\cos\theta_{\rm j}) \right] = \nonumber\\
& = & -x_a p_{{\rm j}T}\sqrt{s} \left[ e^{-\eta_{\rm j}} -
 2\left(\frac{k_{\perp a}}{x_a\sqrt{s}}\right)\cos\phi_a +
   \left(\frac{k_{\perp a}}{x_a\sqrt{s}}\right)^2 e^{\eta_{\rm j}} \right] \,, \\
\hat u  =  (p_b-p_c)^2 & = &  - x_b E_{\rm j}\sqrt{s}\,\left[ 1+\cos\theta_{\rm j} -
2 \left(\frac{k_{\perp b}}{x_b\sqrt{s}}\right) \sin\theta_{\rm j}\cos\phi_b +
\left(\frac{ k_{\perp b}}{x_b\sqrt{s}}\right)^2 (1-\cos\theta_{\rm j}) \right] = \nonumber\\
& = & -x_b p_{{\rm j}T}\sqrt{s} \left[ e^{\eta_{\rm j}} -
 2\left(\frac{k_{\perp b}}{x_b\sqrt{s}}\right)\cos\phi_b +
   \left(\frac{k_{\perp b}}{x_b\sqrt{s}}\right)^2 e^{-\eta_{\rm j}} \right] \,,
\end{eqnarray}
with the condition $\hat s + \hat t + \hat u = 0$ giving an additional constraint.

The helicity frame of the fragmenting parton $c$ has axes denoted by $\hat{\bm{x}}_{\rm j}$ , $\hat{\bm{y}}_{\rm j}$, $\hat{\bm{z}}_{\rm j}$, with $\hat{\bm{z}}_{\rm j}$ along the direction of motion of $c$. It
can be reached  from the hadronic c.m.\ frame by performing a simple rotation
by the angle $\theta_{\rm j}$ around $\hat{\bm{Y}}_{\rm cm}\equiv \hat{\bm{y}}_{\rm j}$,  as can be seen from Fig.\ \ref{fig-kinem}.  Hence, in this frame,
\begin{eqnarray}
\tilde{p}_c &=& \tilde{p}_{\rm j} = E_{\rm j}(1,0,0,1)\nonumber\\
\tilde{p}_\pi &=& \left(E_\pi,\bm{k}_{\perp\pi},\sqrt{E_\pi^2-\bm{k}_{\perp\pi}^2}\,\right) =
\left(E_\pi,{k}_{\perp\pi}\cos\phi_\pi^H,{k}_{\perp\pi}
\sin\phi_\pi^H,\sqrt{E_\pi^2-\bm{k}_{\perp\pi}^2}\,\right)\,,
\label{4mom-H}
\end{eqnarray}
with  $\phi_\pi^H$ being the azimuthal angle of the pion three-momentum around the jet axis, as measured in the fragmenting parton helicity frame. From Eq.\ (\ref{4mom-H}), one can obtain the expression for the light-cone momentum
fraction of the pion,
\begin{equation}
z = \frac{\tilde{p}_\pi^+}{\tilde{p}_c^+}\equiv \frac{\tilde{p}_\pi^+}{\tilde{p}_{\rm j}^+} = \frac{\tilde{p}^0_\pi  +  \tilde{p}^3_\pi}{\tilde{p}^0_{\rm j} + \tilde{p}^3_{\rm j}} =
\frac{E_\pi+\sqrt{E_\pi^2-\bm{k}_{\perp\pi}^2}}{2E_{\rm j}}\, .
\label{z-def}
\end{equation}

By writing down explicitly the three-momentum of the pion $\bm{p}_\pi$
in the parton $c$ helicity frame and in the hadronic c.m.~frame respectively,
\begin{eqnarray}
\bm{p}_\pi &=& k_{\perp\pi}\cos\phi_\pi^H\hat{\bm{x}}_{\rm j}+
k_{\perp\pi}\sin\phi_\pi^H\hat{\bm{y}}_{\rm j}+
\sqrt{E_\pi^2-\bm{k}_{\perp\pi}^2}\,\hat{\bm{z}}_{\rm j}\nonumber\\
&=& \Bigl[\,k_{\perp\pi}\cos\phi_\pi^H\cos\theta_{\rm j}+
\sqrt{E_\pi^2-\bm{k}_{\perp\pi}^2}\sin\theta_{\rm j}\,\Bigr]\hat{\bm{X}}_{\rm cm}+
k_{\perp\pi}\sin\phi_\pi^H\hat{\bm{Y}}_{\rm cm} \label{pi-H-cm}\\
&& \qquad + ~\Bigl[\,-k_{\perp\pi}\cos\phi_\pi^H\sin\theta_{\rm j}+
\sqrt{E_\pi^2-\bm{k}_{\perp\pi}^2}\cos\theta_{\rm j}\,\Bigr]\hat{\bm{Z}}_{\rm cm}\,,\nonumber
\end{eqnarray}
one finds  that the intrinsic transverse momentum of the pion in the hadronic c.m.~frame can be written as
\begin{equation}
\bm{k}_{\perp\pi} = k_{\perp\pi}\cos\phi_\pi^H\cos\theta_{\rm j}\hat{\bm{X}}_{\rm cm}+
k_{\perp\pi}\sin\phi_\pi^H\hat{\bm{Y}}_{\rm cm}-
k_{\perp\pi}\cos\phi_\pi^H\sin\theta_{\rm j}\hat{\bm{Z}}_{\rm cm}\,.
\label{kpi-cm}
\end{equation}
Therefore, denoting by $\phi_{k}$ the azimuthal angle of $\bm{k}_{\perp\pi}$,
\emph{as measured in the hadronic c.m.~frame}, one obtains
\begin{equation}
\tan\phi_{k} = \frac{\tan\phi_\pi^H}{\cos\theta_{\rm j}}~.
\label{tan-phik}
\end{equation}
In Ref.~\cite{Yuan:2007nd}, where only forward jet production was considered,
azimuthal asymmetries were given in terms of  $\phi_k$ (named $\phi_h$ there).
In this kinematical configuration, $\cos\theta_{\rm j} \to 1$ and the angles
 $\phi_\pi^H$ and $\phi_k$ become practically identical. On the other hand, for central-rapidity jets ($\theta_{\rm j}=\pi/2$), $\phi_k=\pi/2$, implying that
 azimuthal asymmetries expressed as function of $\phi_k$ would be artificially
suppressed. For this reason, $\phi_\pi^H$ has to be considered as the
physically relevant angle in the present  analysis.

\section{\label{GPM} The generalized parton model}

The single transversely polarized cross section for the process $p(S) + p\to {\rm jet}+ \pi + X$ has been calculated in the GPM framework, using the helicity
formalism, in Ref.~\cite{D'Alesio:2010am},
to which we refer for further details. Its final expression has the following
general structure,
\begin{eqnarray}
2{\rm d}\sigma(\phi_{S},\phi_\pi^H) &\sim & {\rm d}\sigma_0
+{\rm d}\Delta\sigma_0\sin\phi_{S}+
{\rm d}\sigma_1\cos\phi_\pi^H+ {\rm d}\sigma_2\cos2\phi_\pi^H+
{\rm d}\Delta\sigma_{1}^{-}\sin(\phi_{S}-\phi_\pi^H)
\nonumber\\
&& \qquad +~{\rm d}\Delta\sigma_{1}^{+}\sin(\phi_{S}+\phi_\pi^H)
+{\rm d}\Delta\sigma_{2}^{-}\sin(\phi_{S}-2\phi_\pi^H)+
{\rm d}\Delta\sigma_{2}^{+}\sin(\phi_{S}+2\phi_\pi^H)\,,
\label{d-sig-phi-SA}
\end{eqnarray}
where, as discussed in Section~\ref{kinematics}, $\phi_\pi^H$ is the azimuthal angle of the pion three-momentum around the jet axis and $\phi_S$ is the azimuthal angle of the spin polarization vector $S$ of the polarized proton,
as measured in the hadronic c.m.\ frame. The numerator of the related single spin asymmetry is given by
\begin{eqnarray}
{\rm d}\sigma(\phi_{S},\phi_\pi^H)-
{\rm d}\sigma(\phi_{S}+\pi,\phi_\pi^H)
& \sim &  {\rm d}\Delta\sigma_0\sin\phi_{S}+
{\rm d}\Delta\sigma_{1}^{-}\sin(\phi_{S}-\phi_\pi^H)+
{\rm d}\Delta\sigma_{1}^{+}\sin(\phi_{S}+\phi_\pi^H)\nonumber\\
&&+ \;{\rm d}\Delta\sigma_{2}^{-}\sin(\phi_{S}-2\phi_\pi^H)+
{\rm d}\Delta\sigma_{2}^{+}\sin(\phi_{S}+2\phi_\pi^H)\,,
\label{num-asy-gen}
\end{eqnarray}
while for the denominator we have
\begin{equation}
{\rm d}\sigma(\phi_{S},\phi_\pi^H)+
{\rm d}\sigma(\phi_{S}+\pi,\phi_\pi^H)
 \equiv 2{\rm d}\sigma^{\rm unp}(\phi_\pi^H) \sim
{\rm d}\sigma_0 + {\rm d}\sigma_1\cos\phi_\pi^H+
{\rm d}\sigma_2\cos2\phi_\pi^H\,.
\label{den-asy-gen}
\end{equation}

The various terms contributing to the cross section in Eq.~(\ref{d-sig-phi-SA})
 are explicitly given by convolutions of different TMD parton distribution
and fragmentation functions with hard scattering (polarized) cross sections.
For example, if we keep only the leading
contributions  after integrating over the intrinsic transverse momenta of the
initial partons, the symmetric term in Eq.~(\ref{den-asy-gen}) is given by
\begin{eqnarray}
\de\sigma_0 \,\equiv\, {E_{\rm j}}\,\frac{\de\sigma_0}{\de^3 {\bm p}_{\rm j}\, \de z\, \de^2 {\bm k}_{\perp \pi}}
& = & \frac{2\alpha_s^2}{s} \sum_{a,b,c,d} \int \frac{\de x_a}{x_a}\,\de^2\bm{k}_{\perp a}
\int \frac{\de x_b}{x_b}\,\de^2{\bm k}_{\perp b} \,
\delta(\hat s+\hat t+\hat u)\,
H^U_{ab\to cd}(\hat s,\hat t,\hat u) \nonumber \\
&&\quad \,\times   f_{a/A}(x_a, {\bm k}_{\perp a}^2) \, f_{b/B}(x_b, {\bm k}_{\perp b}^2)\, D_{1}^c(z, {\bm k}_{\perp \pi}^2)\,,
\label{unp}
\end{eqnarray}
where $H^U_{ab\to cd}(\hat s,\hat t, \hat u)$ is the unpolarized squared hard scattering amplitude for the partonic process  $a\, b\to c\, d$, related
to the elementary cross section as follows:
\begin{equation}
\frac{{\de\hat\sigma}_{ab\to cd}}{\de\hat t} = \frac{\pi\alpha_s^2}{\hat s^2}\,
H^U_{ab\to cd}~.
\end{equation}
 By $f_{a/A}(x_a, {\bm k}_{\perp a}^2)$ and  $f_{b/B}(x_b, {\bm k}_{\perp b}^2)$ we denote the unpolarized TMD distributions for parton $a$ inside hadron $A$ and for parton $b$ inside
hadron $B$, respectively, while  $D_{1}^c(z,\bm{k}_{\perp\pi}^2)$  is
the unintegrated fragmentation function for the unpolarized parton $c$ that
fragments into a pion. The term containing the $\sin\phi_S$ modulation in Eq.~(\ref{num-asy-gen})
is related to the Sivers effect,
\begin{eqnarray}
\de\Delta\sigma_0\sin\phi_S \,\equiv\, {E_{\rm j}}\,\frac{\de\Delta\sigma^{(\rm{Sivers})}}
{\de^3 {\bm p}_{\rm j}\, \de z\, \de^2 {\bm k}_{\perp \pi}}
& = & \frac{2\alpha_s^2}{s} \sum_{a,b,c,d} \int \frac{\de x_a}{x_a}\,\de^2\bm{k}_{\perp a}
\int \frac{\de x_b}{x_b}\,\de^2{\bm k}_{\perp b} \,
\delta(\hat s+\hat t+\hat u)\,
H^U_{ab\to cd}(\hat s,\hat t,\hat u) \nonumber \\
&&\quad \,\times  \Big ( -\frac{k_{\perp a}} {M} \Big ) f_{1T}^{\perp a}(x_a, {\bm k}_{\perp a}^2) \cos\phi_a
\, f_{b/B}(x_b, {\bm k}_{\perp b}^2)\,
D_{1}^c(z, {\bm k}_{\perp \pi}^2)\sin\phi_S\,,
\label{sivers}
\end{eqnarray}
where $M$ is the proton mass and $f_{1T}^{\perp a}(x_a, \bm{k}_{\perp a}^2)$
 the Sivers function, also denoted as
 $\Delta^{N}\!f_{a/p^\uparrow}=-2(k_\perp/M)f_{1T}^{\perp a}$~\cite{Bacchetta:2004jz}.
Notice that, for a direct comparison with the CGI~GPM approach,
in this review we adopt the so-called Amsterdam notation~\cite{Mulders:1995dh,Boer:1997nt}
instead of the usual GPM notation~\cite{Anselmino:2005sh,D'Alesio:2007jt}.

The term containing the $\sin(\phi_S-\phi_\pi^H)$ modulation in Eq.~(\ref{num-asy-gen}) corresponds
to the Collins effect,
\begin{eqnarray}
\de\Delta\sigma_1^-\sin(\phi_S-\phi_\pi^H) & \equiv &  {E_{\rm j}}\,\frac{\de\Delta\sigma^{(\rm{Collins})}}
{\de^3 {\bm p}_{\rm j}\, \de z\, \de^2 {\bm k}_{\perp \pi}}
\, = \, \frac{2\alpha_s^2}{s} \sum_{a,b,c,d} \int \frac{\de x_a}{x_a}\,\de^2\bm{k}_{\perp a}
\int \frac{\de x_b}{x_b}\,\de^2{\bm k}_{\perp b} \,
\delta(\hat s+\hat t+\hat u)\,
H^U_{ab\to cd}(\hat s,\hat t,\hat u) \nonumber \\
& & \times\,   h_{1}^a(x_a, {\bm k}_{\perp a}^2) \cos(\phi_a-\psi)\, f_{b/B}(x_b, {\bm k}_{\perp b}^2)
\frac{k_{\perp \pi}}{z M_\pi}H_{1}^{\perp c}(z, {\bm k}_{\perp \pi}^2)
\,d_{NN}(\hat s,\hat t,\hat u) \,
\sin(\phi_{S}-\phi_\pi^H)\,,
\label{collins}
\end{eqnarray}
where the Collins fragmentation function of the struck quark $c$,
$H_{1}^{\perp c}(z,\bm{k}_{\perp\pi}^2)$
(or $\Delta^N\!D_{\pi/c^\uparrow}=2(k_{\perp\pi}/zM_\pi)H_{1}^{\perp c})$,
is convoluted with the unintegrated
transversity  distribution, $h_1^a(x_a, \bm{k}_{\perp a}^2)$, that is the
distribution of transversely polarized quarks in a transversely polarized
hadron. In Eq.\ (\ref{collins}), $M_\pi$ is the pion mass,
$d_{NN}$ is the spin transfer asymmetry for the partonic process
$a^\uparrow b\to c^\uparrow d$,
\begin{equation}
d_{NN} =  \frac{\sigma^{a^\uparrow b\to c^\uparrow d} - \sigma^{a^\uparrow b\to c^\downarrow d}}{\sigma^{a^\uparrow b\to c^\uparrow d} + \sigma^{a^\uparrow b\to c^\downarrow d}} \,,
\end{equation}
and $\psi$ the corresponding  azimuthal phase~\cite{D'Alesio:2010am}.

In order to single out the different contributions to the polarized cross section, we introduce the following average values of the circular
functions of $\phi_{S}$ and $\phi_\pi^H$ appearing in Eq.~(\ref{d-sig-phi-SA}),
\begin{equation}
\langle\,W(\phi_{S},\phi_\pi^H)\,\rangle(\bm{p}_{\rm j},z,k_{\perp\pi})=
\frac{\int{\rm d}\phi_{S}\,{\rm d}\phi_\pi^H\,
W(\phi_{S},\phi_\pi^H)\,{\rm d}\sigma(\phi_{S},\phi_\pi^H)}
{\int{\rm d}\phi_{S}\,{\rm d}\phi_\pi^H{\rm\, d}\sigma(\phi_{S},\phi_\pi^H)}\,.
\label{average}
\end{equation}
For single spin asymmetries one can preferably define azimuthal moments,
similarly to the SIDIS case,
\begin{eqnarray}
A_N^{W(\phi_{S},\phi_\pi^H)}(\bm{p}_{\rm j},z,k_{\perp\pi})
&=&
2\,\frac{\int{\rm d}\phi_{S}\,{\rm d}\phi_\pi^H\,
W(\phi_{S},\phi_\pi^H)\,[{\rm d}\sigma(\phi_{S},\phi_\pi^H)-
{\rm d}\sigma(\phi_{S}+\pi,\phi_\pi^H)]}
{\int{\rm d}\phi_{S}\,{\rm d}\phi_\pi^H\,
[{\rm d}\sigma(\phi_{S},\phi_\pi^H)+
{\rm d}\sigma(\phi_{S}+\pi,\phi_\pi^H)]}\,,
\label{gen-mom}
\end{eqnarray}
with $W(\phi_{S},\phi_\pi^H)$ now being one of the angular modulations in
Eq.\ (\ref{num-asy-gen}). In the following we will focus mainly on
the two observables that are the most relevant from the phenomenological point of view: the Collins and the Sivers contributions to $A_N$, namely
$A_N^{\sin(\phi_S-\phi_\pi^H)}$ and   $A_N^{\sin\phi_S}$.

\section{Phenomenological results}

Here as well as in the following sections we review some phenomenological
implications of the TMD generalized parton model approach
for the $p^\uparrow p\to {\rm jet}\,\pi\,X$ and $p^\uparrow p\to {\rm jet}\,X$ processes
in kinematical configurations accessible at RHIC by the STAR and PHENIX experiments.
We consider both central ($\eta_{\rm j}=0$) and forward
($\eta_{\rm j}=3.3$) (pseudo)rapidity configurations,
at c.m.~energies $\sqrt{s} =$ 200 GeV and 500 GeV.
A more detailed account and additional phenomenological results are given in Ref.~\cite{D'Alesio:2010am}.

Preliminary STAR results at $\sqrt{s}=200$ GeV for the Collins azimuthal asymmetry in the process
$p^\uparrow p \to {\rm jet}\,\pi^{\pm}\,X$ in the mid-rapidity region~\cite{Fatemi:2012ry}
and for the Collins and Sivers azimuthal asymmetries in $p^\uparrow p \to {\rm jet}\,\pi^{0}\,X$
at forward rapidities~\cite{Poljak:2011vu} are also available.
A phenomenological analysis of these results in the GPM approach, with proper account of all
jet kinematical cuts, is in progress and will be presented elsewhere~\cite{dalesio:2013prg}.

In the sequel TMD parton distribution and
fragmentation functions are parameterized with a simplified
functional dependence on
the parton light-cone momentum fraction and on the transverse
motion, which are completely factorized.
Notice however that kinematical constraints due to usual
parton model requirements (implemented in numerical calculations)
effectively lead to correlations between the light-cone
momentum fraction and the transverse momentum, particularly at
very small and very large ($\to 1$) momentum fractions
(for more details, see e.g.~appendix A of Ref.~\cite{D'Alesio:2004up}).
 Moreover, we assume a Gaussian-like flavour-independent
shape for the transverse momentum component.
Preliminary lattice QCD calculations seem to support the validity of this
assumption, see e.g.~Ref.~\cite{Hagler:2009ni}.

Concerning the parameterizations of the quark transversity and
Sivers distributions, and of the quark Collins functions,
we will consider two sets:
SIDIS~1 \cite{Anselmino:2005ea, Anselmino:2007fs} and
SIDIS~2 \cite{Anselmino:2008sga,Anselmino:2008jk}.

The set SIDIS~1 includes the $u$, $d$ quark Sivers functions of Ref.~\cite{Anselmino:2005ea},
the $u$, $d$ quark transversity distributions and the favoured
and disfavoured Collins FFs of Ref.~\cite{Anselmino:2007fs}.
The Kretzer
set~\cite{Kretzer:2000yf} for collinear pion FFs was used.

Instead, the set SIDIS~2 includes the $u$, $d$, and sea-quark
Sivers functions of Ref.~\cite{Anselmino:2008sga} and the updated set
of the $u$, $d$ quark transversity distributions and of the favoured
and disfavoured Collins FFs of Ref.~\cite{Anselmino:2008jk}.
In this case, the DSS set~\cite{deFlorian:2007aj} for collinear pion and kaon FFs was adopted.

In both cases, for the usual collinear parton distributions, the LO
unpolarized set GRV98~\cite{Gluck:1998xa} and the corresponding
longitudinally polarized set GRSV2000~\cite{Gluck:2000dy} (needed in order to
implement the Soffer bound~\cite{Soffer:1994ww} for the transversity distribution)  were adopted.

Notice that quite recently, updated parameterizations of the transversity distribution
and of the Collins function within the GPM approach have been released~\cite{Anselmino:2013vqa}.
Since they are qualitatively similar to those adopted in Ref.~\cite{D'Alesio:2010am},
for ease of comparison they will not be used in the following.

Since the jet transverse momentum
(the hard scale in the process) covers a significant range,
one should properly take into account the QCD evolution
of all TMDs.
On the other hand, a formal proof of TMD factorization for such processes
is still missing and the study of TMD evolution is at present in its earlier stage.
Therefore, we tentatively take into account proper evolution
with scale, at leading order, for the usual collinear PDFs and FFs,
while keeping the transverse momentum component of all TMDs
fixed.

The study of the formal aspects and the related phenomenology of the correct
QCD evolution with scale of TMD PDFs and FFs has received a lot of attention
quite recently. Several papers have investigated proper TMD evolution equations for the Sivers
function and their phenomenological implications, see e.g.~Refs.~\cite{Aybat:2011zv,Aybat:2011ge,Aybat:2011ta,Anselmino:2012aa,Boer:2013zca,Sun:2013dya,Echevarria:2012pw}.
The TMD evolution of the helicity and transversity parton distributions has been
considered e.g.~in Ref.~\cite{Bacchetta:2013pqa}.
No information is available yet on the TMD evolution of the Collins fragmentation
functions.

In all cases considered, $\bm{k}_{\perp\pi}$ is integrated out and, since we are interested in leading particles inside the jet, we  present results obtained
integrating over the light-cone momentum fraction of the observed hadron, $z$,
in the range $z\geq 0.3$. Different choices, according to the kinematical cuts of interest in specific experiments,
can be easily implemented in the numerical calculations.

We have considered first, for $\pi^+$ production only,
an extreme scenario in which the effects of
all TMD functions are over-maximized. By this we mean that all TMDs
are maximized in size by imposing natural positivity bounds.
The transversity distribution has been fixed at the initial scale
by saturating the Soffer bound~\cite{Soffer:1994ww} and then we let it
evolve. Moreover, the relative signs of
all active partonic contributions are chosen so that they
sum up additively. In this way we set
an upper bound on the absolute value of any of the effects playing
a potential role in the azimuthal asymmetries.
Therefore, all effects that are negligible or even
marginal in this scenario may be directly discarded in subsequent
refined phenomenological analyses. See Ref.~\cite{D'Alesio:2010am} for a more detailed discussion.

As a second step in our study we consider, for both neutral and charged pions,
only the dominant contributions, that is the Collins and the Sivers effects,
involving TMD functions for which parameterizations
are available from independent fits to other spin and azimuthal
asymmetries data in SIDIS and $e^+e^-$ processes (the SIDIS~1
and SIDIS~2 sets discussed above).

\subsection{The Collins  asymmetries \label{sec:res-coll}}

The Collins fragmentation function  contributes to two of the
azimuthal moments defined in Eq.\ (\ref{gen-mom}), namely $A_N^ {\sin(\phi_S+\phi_\pi^H)}$ and  $A_N^ {\sin(\phi_S-\phi_\pi^H)}$. In $A_N^ {\sin(\phi_S+\phi_\pi^H)}$ it is convoluted with two different terms:
\begin{equation}
A_N^ {\sin(\phi_S+\phi_\pi^H)}\sim
\left [ h_{1 T}^{\perp q}(x_a,\bm{k}_{\perp a}^2)\otimes f_1(x_b,\bm{k}_{\perp b}^2) +
 f_{1 T}^{\perp}(x_a,\bm{k}_{\perp a}^2) \otimes h_1^{\perp q}(x_b,\bm{k}_{\perp b}^2) \right ]
  \otimes H_1^{\perp q}(z,\bm{k}_{\perp \pi}^2)~.
\label{eq:Coll1}
\end{equation}

The first term is related to the so-called pretzelosity distribution $h_{1 T}^{\perp q}$, while the second one, which enters also in the expression for   $A_N^ {\sin(\phi_S-\phi_\pi^H)}$, involves in the convolution the Sivers and
Boer-Mulders ($h_1^{\perp q}$) functions.
\begin{figure}[t]
\begin{center}
 \includegraphics[angle=0,width=0.4\textwidth]{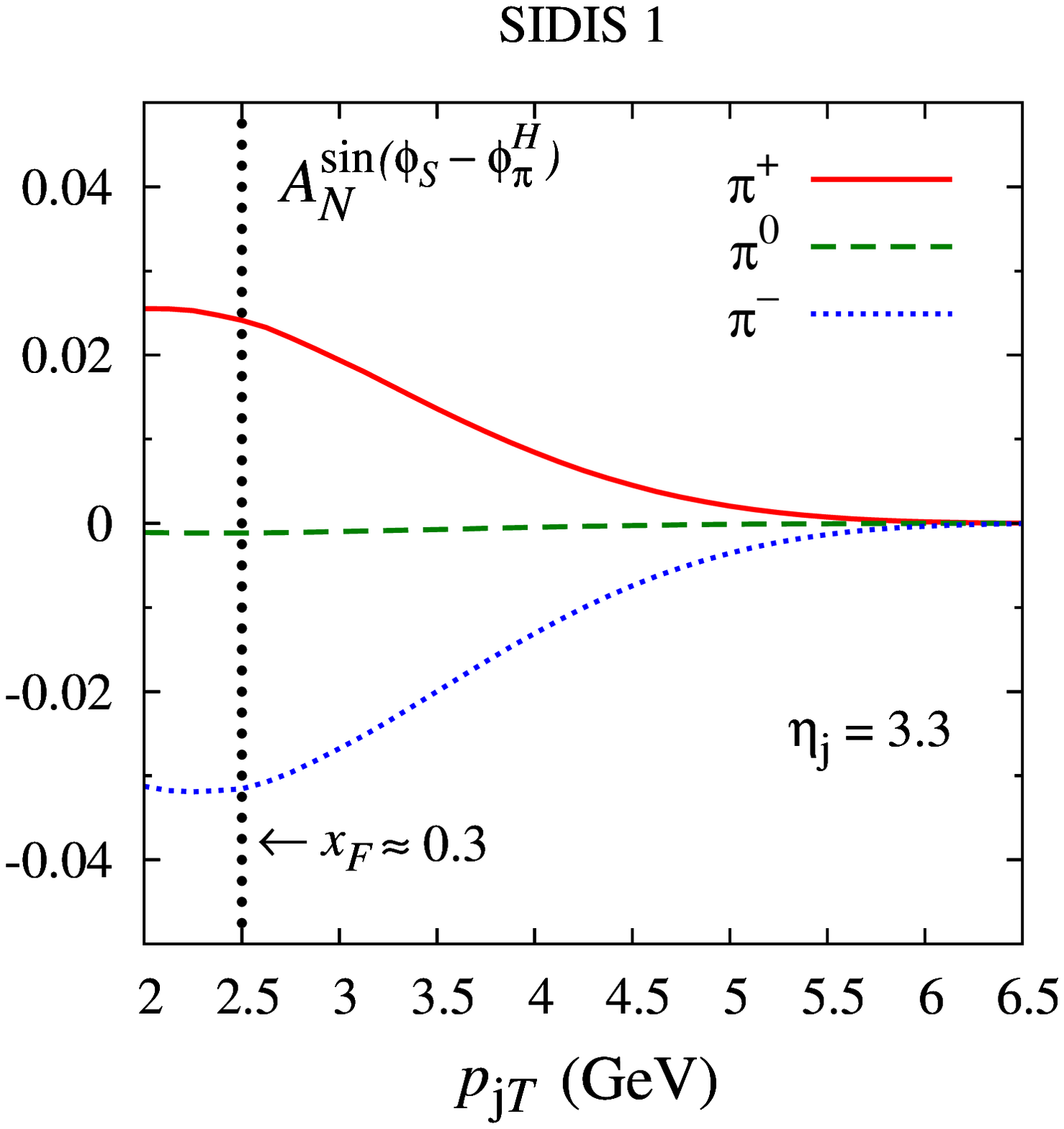}
 \includegraphics[angle=0,width=0.4\textwidth]{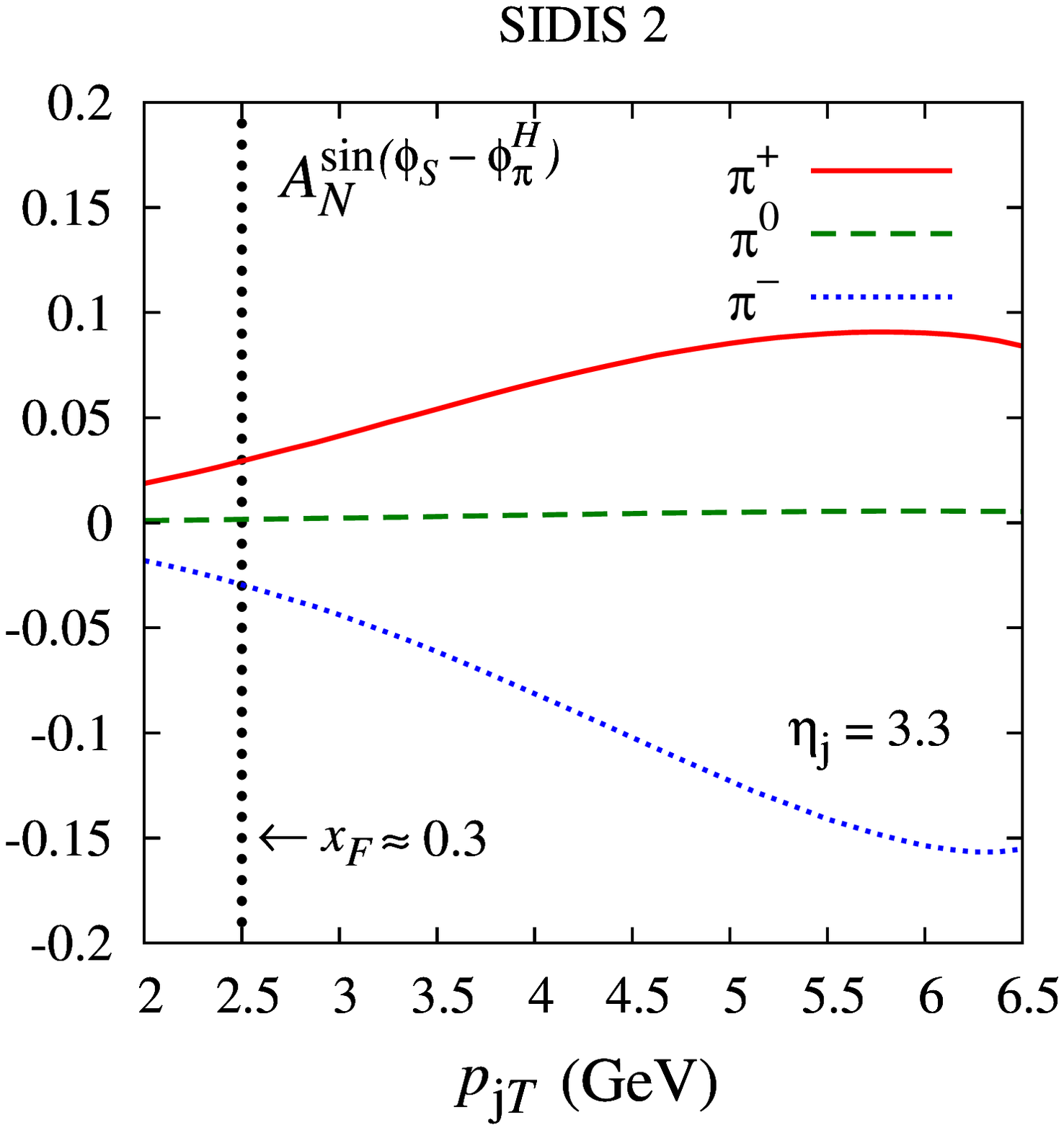}
 \caption{The Collins asymmetry  $A_N^{\sin(\phi_{S}-\phi_\pi^H)}$  for the process $p^\uparrow \, p\to {\rm jet}\,
\pi \, X$, as a function of $p_{{\rm j} T}$, at fixed value of the rapidity $\eta_{\rm j}$ and c.m.~energy
$\sqrt{s}= 200$ GeV. Estimates are obtained by adopting the parameterizations
SIDIS~1 (left panel)
and SIDIS~2 (right panel). The dotted vertical line delimits the region $x_F \approx 0.3$,
beyond which the currently available parameterizations for the
quark transversity distributions, extracted from SIDIS data, are affected
by large uncertainties.
\label{asy-an-coll-par200} }
\end{center}
\end{figure}
\begin{figure}[t]
\begin{center}
 \includegraphics[angle=0,width=0.4\textwidth]{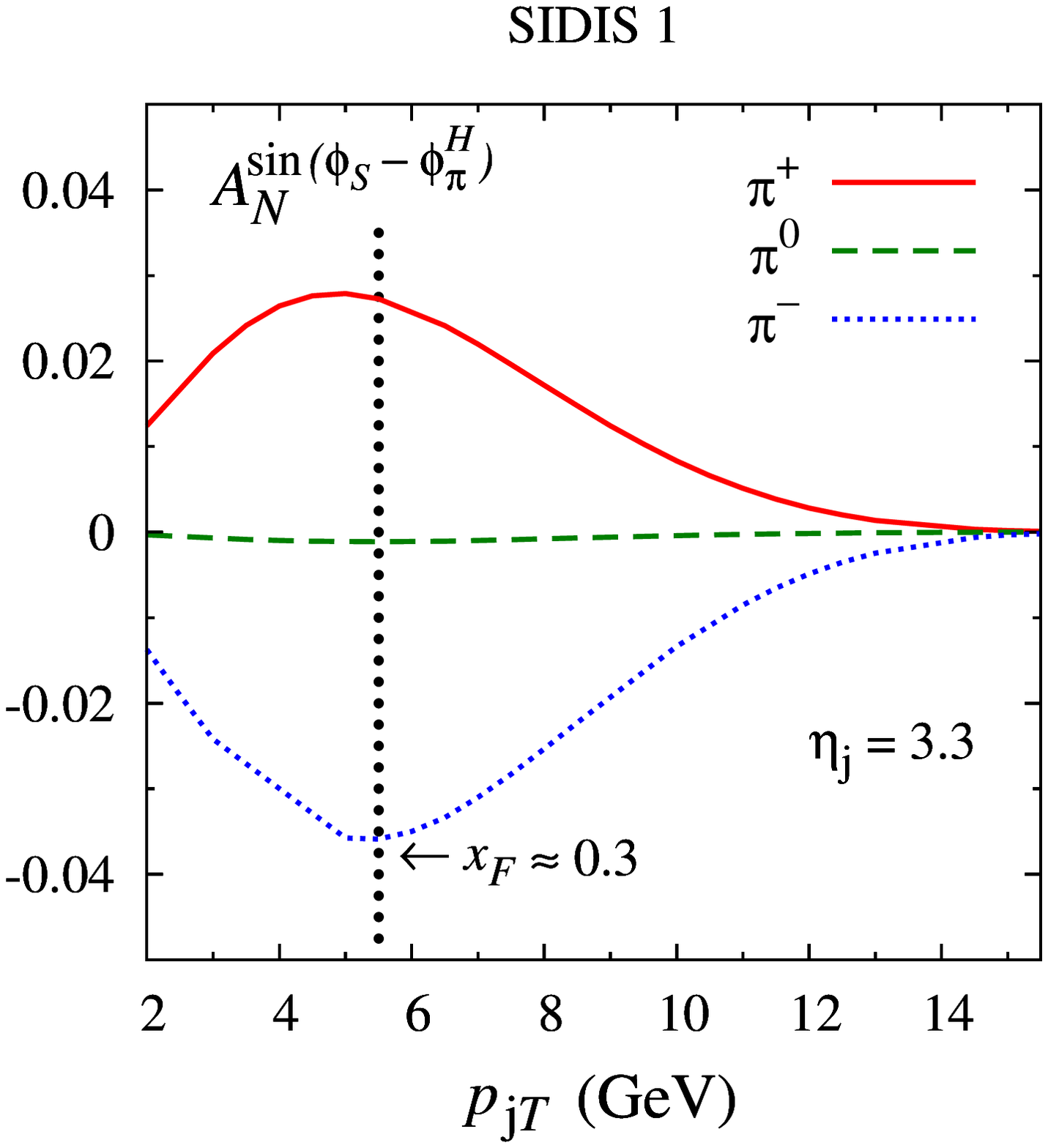}
 \includegraphics[angle=0,width=0.4\textwidth]{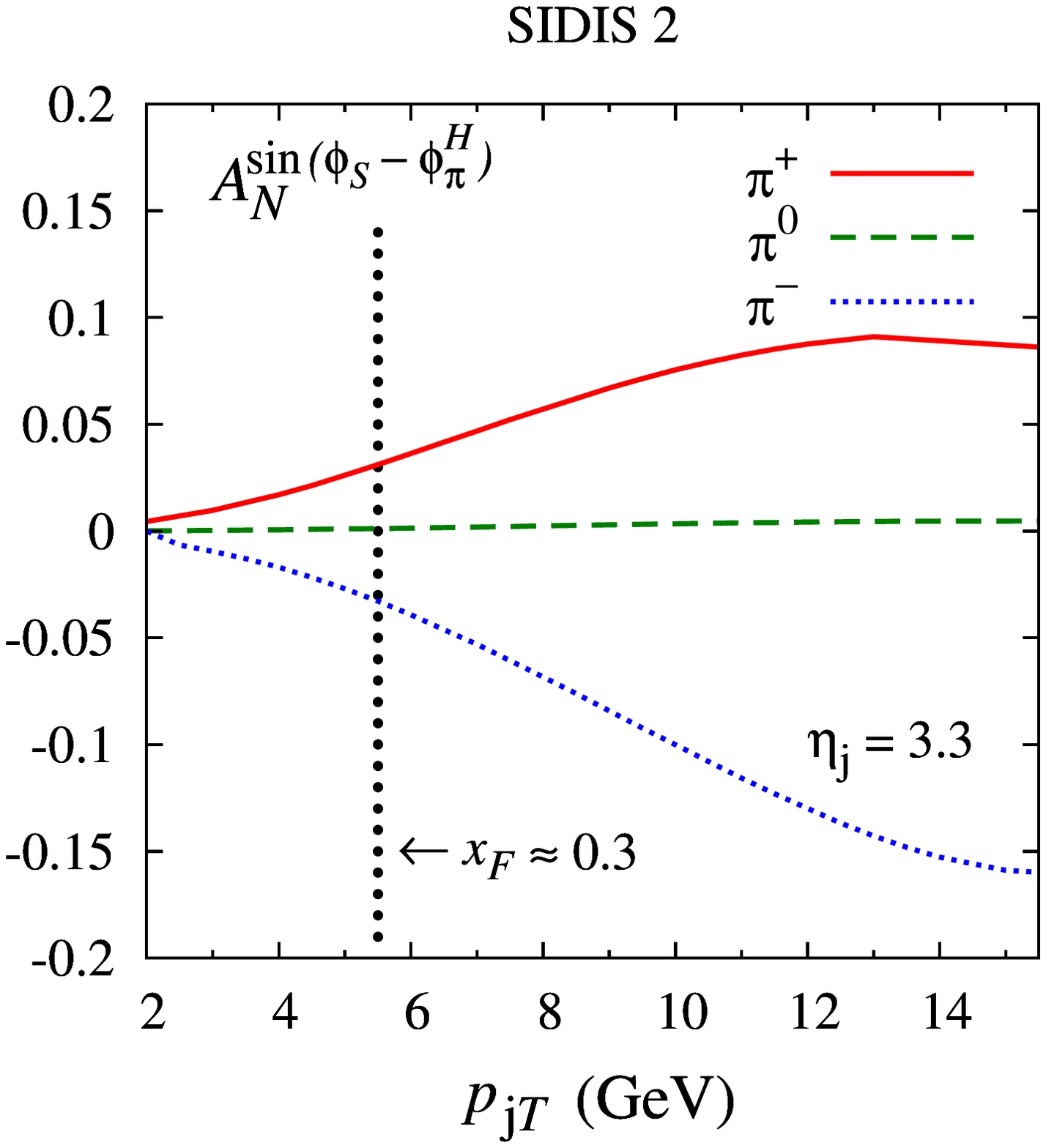}
 \caption{The same as for Fig.~\ref{asy-an-coll-par200}, but at c.m.~energy
 $\sqrt{s}= 500$ GeV.
\label{asy-an-coll-par500} }
\end{center}
\end{figure}
As described above, and in more detail in Ref.~\cite{D'Alesio:2010am},
it has been checked  that the upper bound of this asymmetry is always negligible, hence it will not
be considered again in the following. Same conclusions hold for the
Collins-like azimuthal moment $ A_N^ {\sin(\phi_S+2 \phi_\pi^H)}$
originating from the fragmentation of linearly polarized gluons, which has a
structure similar to Eq.~(\ref{eq:Coll1}), with quarks replaced by
gluons.

The azimuthal asymmetry $A_N^{\sin(\phi_{S}- \phi_\pi^H)}$ is
dominated  by a  convolution of the transversity distribution and the Collins
 fragmentation function,
\begin{equation}
A_N^{\sin(\phi_S-\phi_\pi^H)} \sim h_1^q(x_a,\bm{k}_{\perp a}^2) \otimes
 f_1(x_b,\bm{k}_{\perp b}^2) \otimes H_1^{\perp\, q}(z,\bm{k}_{\perp \pi}^2)\,,
\label{eq:Coll2}
\end{equation}
see  Eq.\ (\ref{collins}).
A similar expression holds for its gluonic counterpart $A_N^{\sin(\phi_{S}- 2 \phi_\pi^H)}$. Their upper bounds turn out to be sizeable, at least  in some kinematic domains \cite{D'Alesio:2010am}.

In Figs.~\ref{asy-an-coll-par200} and~\ref{asy-an-coll-par500} we show our
estimates for $A_N^{\sin(\phi_{S}- \phi_\pi^H)}$ at the RHIC energies $\sqrt{s} = 200$ GeV
and 500 GeV respectively, as a function of the transverse momentum of the jet, $p_{{\rm j}T}$, and at fixed jet rapidity ($\eta_{{\rm j}}$= 3.3). These results have been obtained by adopting the parameterizations SIDIS~1 and  SIDIS~2.
Notice that while the results of Fig.~\ref{asy-an-coll-par200} are taken from Ref.~\cite{D'Alesio:2010am},
those of Fig.~\ref{asy-an-coll-par500} are presented here for the first time.
Our prediction of an almost vanishing asymmetry for neutral
pions, confirmed very recently by preliminary data at $\sqrt{s} = 200$ GeV from the STAR Collaboration~\cite{Poljak:2011vu},
is a consequence of the comparable size and the opposite sign, in both parameterizations, of the favoured (e.g.~$u\to\pi^+$) and disfavoured (e.g.~$d\to\pi^+$) Collins fragmentation functions. In fact, because of isospin invariance,
the Collins function for neutral pions is given by half the sum
of the fragmentation functions for charged pions, hence turning out to be
very small.
In addition, further cancellations among quark contributions are due to the
 relative opposite sign of the  transversity distribution for $u$ and $d$
flavours. Concerning charged pions, the two parameterizations give comparable
results only in the kinematic domain where the Feynman variable
$x_F=2 p_{{\rm j} L}/\sqrt{s}$ is equal to or smaller than the value
$x_F \approx 0.3$, denoted by the dotted vertical lines in
  Figs.~\ref{asy-an-coll-par200},~\ref{asy-an-coll-par500} (notice the different scales used in the
 two panels). This corresponds to the  Bjorken $x$ region covered by
the SIDIS data that have been used to determine the available
parameterizations for the transversity distributions. Extrapolation beyond
$x_F \approx 0.3$, where transversity is  not constrained,
leads to completely different estimates at large $p_{{\rm j} T}$,
as shown in the figures.

Based on these considerations, in a recent paper~\cite{Anselmino:2012rq}
(to which we refer for more details) a different and complementary analysis
(denoted as ``scan procedure") has been performed.
The large $x$ behaviour of the quark transversity distribution
is mainly controlled by the parameters $\beta_q$ ($q=u,\, d$) in the factor
$(1-x)^{\beta_q}$ of the parametrization~\cite{D'Alesio:2010am}, which are basically unconstrained
by SIDIS data. Therefore, starting from a reference fit (with a given total $\chi^2$, $\chi^2_0$)
to updated SIDIS and $e^+e^-$ data (hence, although using the same
collinear PDFs and FFs, slightly different from the SIDIS~1 set)
the following procedure has been implemented:
First, we fix $\beta_{u,d}$ within the range $[0,4]$ by discrete steps of $0.5$, for a total of 81
different $\{\beta_u,\beta_d\}$ configurations;
secondly, for each of these  $\{\beta_u,\beta_d\}$ pairs, we perform a new fit of the
other parameters and evaluate its corresponding total $\chi^2$.
Only those configurations with a $\Delta\chi^2=\chi^2-\chi^2_0$ less than
a statistically significant reference value (see Ref.~\cite{Anselmino:2012rq} for further details)
have been kept. In practice, in this case all 81 configurations fulfill the selection criterium, reinforcing the conclusion
that presently available SIDIS data do not constrain the large $x$
behaviour of the TMD transversity distribution.

For a given process of interest and the related azimuthal asymmetries,
like e.g.~the inclusive particle production in polarized $pp$ collisions
studied in this review (in particular in the large $x_F$ region),
the final step of the scan procedure consists in taking the full envelope of the values
of the asymmetry generated by considering all the selected configuration sets.
This envelope gives an estimate of the uncertainty in the asymmetry calculation
due to the limited $x_B$ range covered by SIDIS data and the consequent indeterminacy
in the large $x$ behaviour of the quark transversity distribution.

\begin{figure}[t]
\begin{center}
 \includegraphics[angle=0,width=0.4\textwidth]{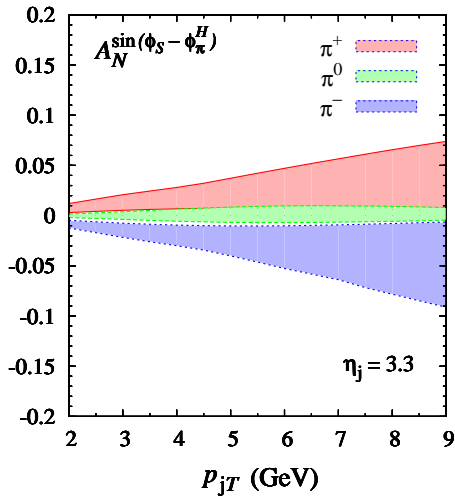}
 \caption{
 Scan bands (that is, the envelope of possible values) for the
 Collins azimuthal asymmetry
 $A_N^{\sin(\phi_{S}-\phi_\pi^H)}$  for the process $p^\uparrow \, p\to {\rm jet}\,
 \pi \, X$, as a function of $p_{{\rm j} T}$, at fixed value of the pseudorapidity,
  $\eta_{\rm j}=3.3$ and c.m.~energy $\sqrt{s}= 500$ GeV.
  The shaded bands are generated following the scan procedure explained in the text
  (see Ref.~\cite{Anselmino:2012rq} for more details).
\label{asy-an-coll-scan-500} }
\end{center}
\end{figure}

As an example, in Fig.~\ref{asy-an-coll-scan-500} we show the resulting scan bands for the
Collins azimuthal asymmetry $A_N^{\sin(\phi_{S}- \phi_\pi^H)}$ for neutral and
charged pions at the RHIC c.m.~energy $\sqrt{s}=500$ GeV, as a function of the jet
transverse momentum and fixed jet pseudorapidity, $\eta_{\rm j}=3.3$ (that is, the same
kinematical configuration of Fig.~\ref{asy-an-coll-par500}).

It is clear from this plot how the uncertainty on the asymmetry grows as $p_{{\rm j} T}$
(and consequently $x_F$) increases. This information is complementary and integrates
the indications obtained comparing the results of the specific SIDIS~1 and SIDIS~2 sets in
Figs.~\ref{asy-an-coll-par200},~\ref{asy-an-coll-par500}.

It is also clear that future measurements of the Collins
asymmetries for charged pions in $p^\uparrow p \to {\rm jet}\, \pi\, X$ processes
would be very helpful in delineating
the large $x$ behaviour of the quark transversity distributions.
We point out that in the central rapidity region these asymmetries are much
smaller. Nevertheless, they are currently under
active investigation by the STAR Collaboration \cite{Poljak:2011vu,Fatemi:2012ry}.

Finally, analogous estimates for the azimuthal moment
$A_N^{\sin(\phi_{S}- 2 \phi_\pi^H)}$  cannot be provided, since
the underlying TMD gluon distribution and fragmentation functions
 are still completely unknown.

\subsection{The Sivers asymmetries}

\begin{figure}[t]
\begin{center}
 \includegraphics[angle=0,width=0.35\textwidth]{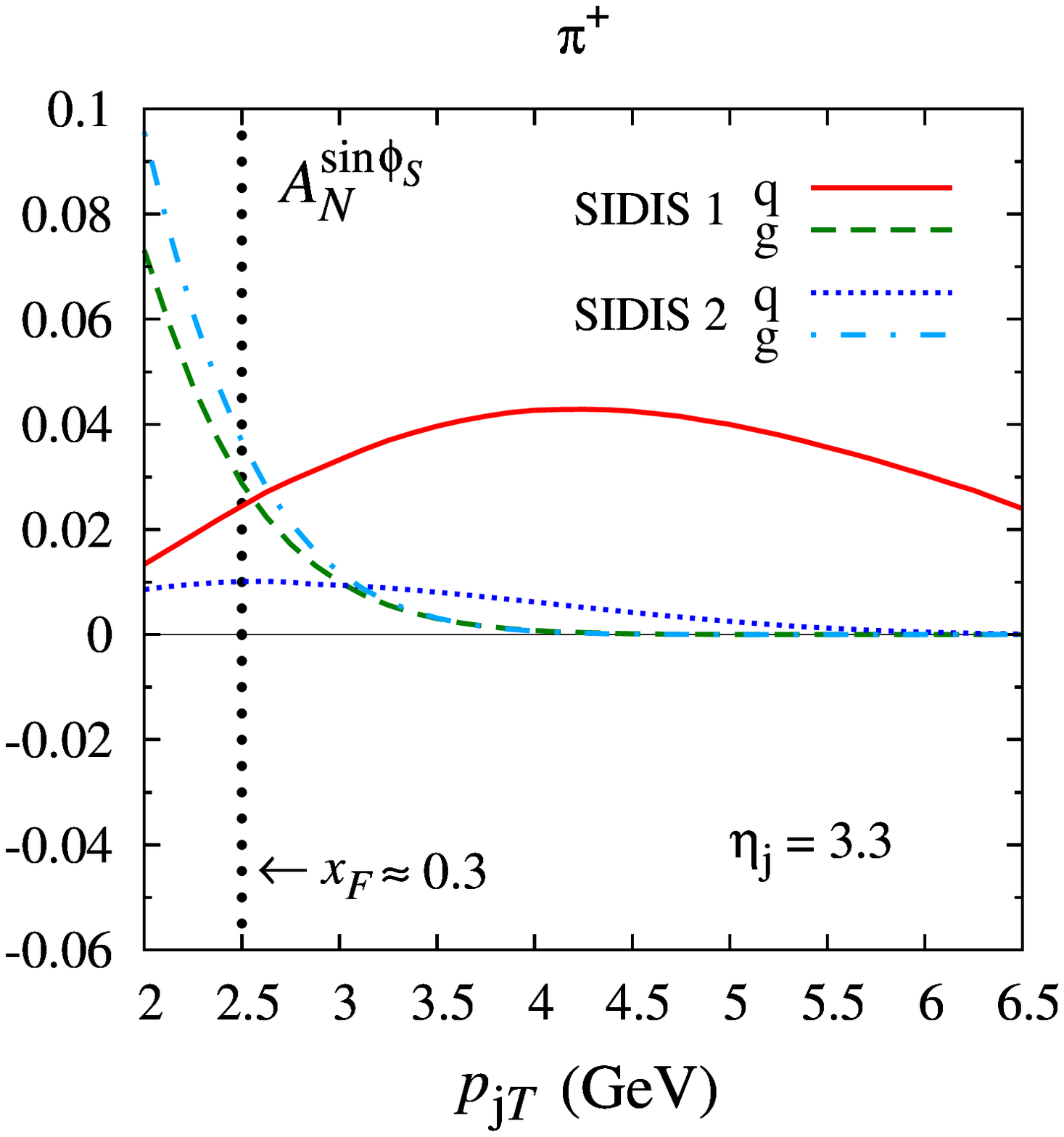}
 \hspace*{-20pt}
 \includegraphics[angle=0,width=0.35\textwidth]{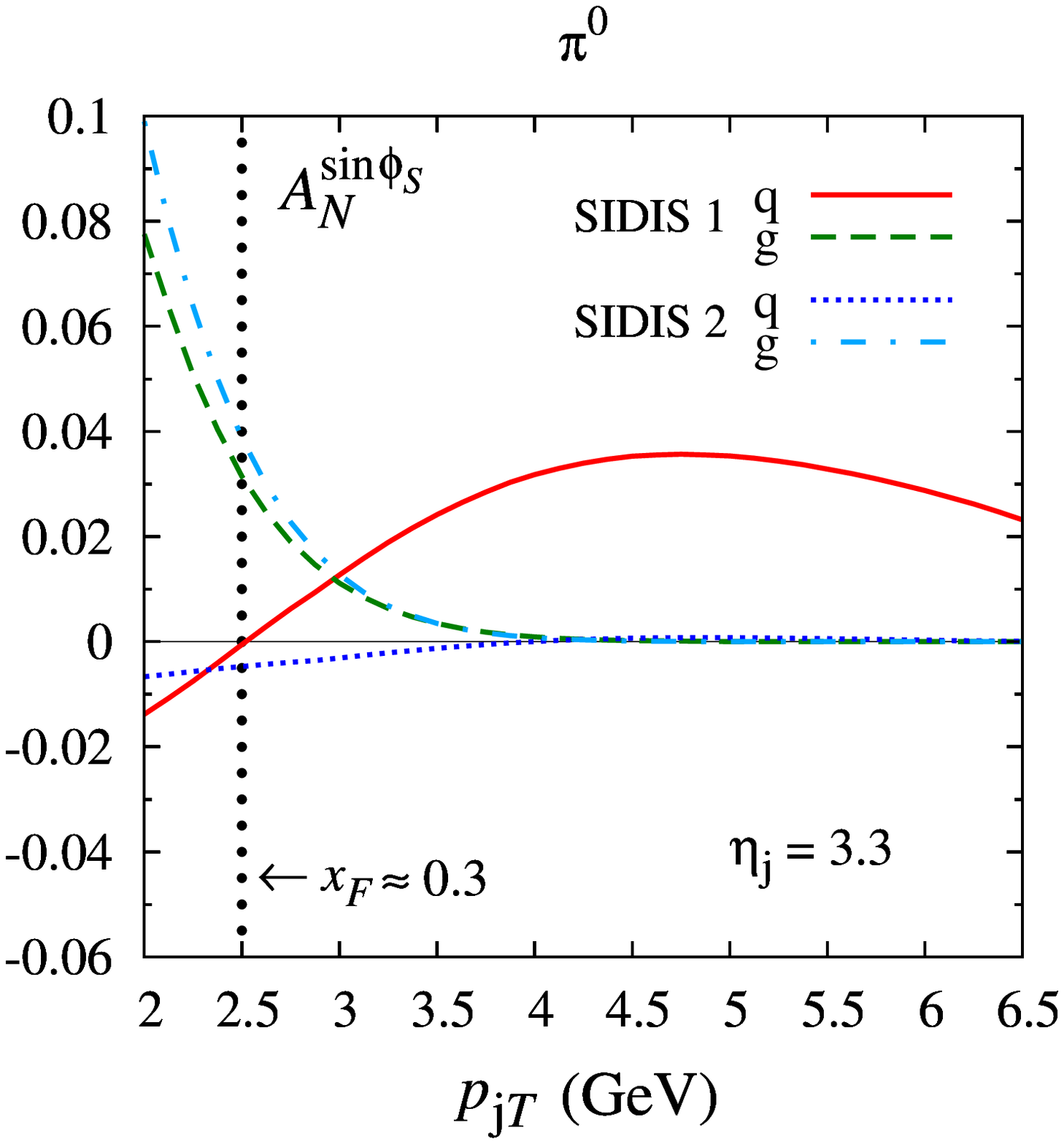}
 \hspace*{-20pt}
 \includegraphics[angle=0,width=0.35\textwidth]{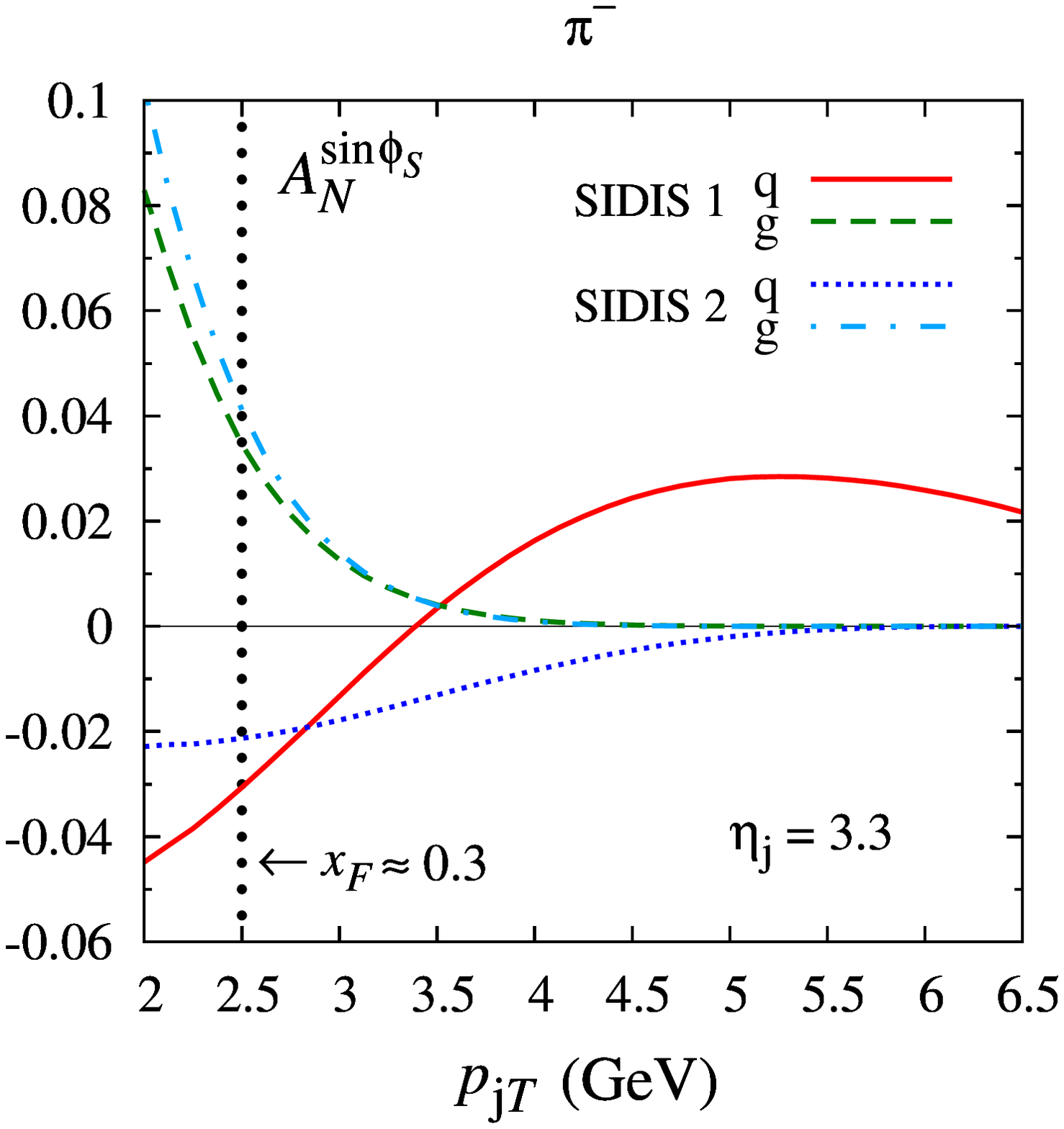}
 \caption{The Sivers asymmetry  $A_N^{\sin\phi_{S}}$
 for the process $p^\uparrow \, p\to {\rm jet}\, \pi \, X$, as a function
of $p_{{\rm j} T}$, at fixed value of the rapidity $\eta_{\rm j}$ and c.m.\ energy  $\sqrt{s}= 200$ GeV. Estimates for the quark contribution are obtained
by adopting the
parametrization sets SIDIS~1 and SIDIS~2. The gluon Sivers function is
assumed to be positive and to saturate an updated version of the
bound in Ref.~\cite{Anselmino:2006yq}. The dotted vertical line delimits the region $x_F\approx 0.3$, beyond which  the currently available
parameterizations for the
quark Sivers function, extracted from SIDIS data, are affected by large
uncertainties.
 \label{asy-an-siv-par200} }
\end{center}
\end{figure}

In analogy to Eqs.\ (\ref{eq:Coll1}) and (\ref{eq:Coll2}), the azimuthal
moment $A_N^ {\sin\phi_S}$ can be written
schematically as
\begin{equation}
A_N^ {\sin\phi_S}\sim  f_{1 T}^{\perp}(x_a,\bm{k}_{\perp a}^2) \otimes
 f_1(x_b,\bm{k}_{\perp b}^2)  \otimes D_1(z,\bm{k}_{\perp \pi}^2)\,,
\label{eq:Siv}
\end{equation}
{\it i.e.}\ as a convolution of the Sivers function for the parton inside
the transversely polarized proton with the unpolarized TMD
distribution and fragmentation functions of the two other active partons
in the hard scattering.
The explicit expression for the
numerator of the asymmetry is given in Eq.~(\ref{sivers}). Both the quark and
gluon Sivers functions contribute to this observable, and in principle
these contributions cannot be separated. Nevertheless it should be possible to select between
the two terms by looking at particular kinematic domains in which only
one of them is expected to be sizeable and dominate the asymmetry~\cite{D'Alesio:2010am}.

In Fig.~\ref{asy-an-siv-par200} $A_N^{\sin\phi_{S}}$ is presented,
for both neutral and charged pions, at the c.m.~energy $\sqrt{s}=200$ GeV and
in the forward rapidity region
($\eta_{\rm j}=3.3$), as a function of $p_{{\rm j} T}$.
The quark Sivers contribution is estimated adopting the SIDIS~1 and
SIDIS~2 parameterizations, which give comparable results only in the $p_{{\rm j} T}$ region where they are constrained by SIDIS data (see, as for the case of
the Collins asymmetry, the dotted vertical line). The almost unknown gluon
Sivers function is tentatively taken positive and saturates an updated version
of the bound calculated in Ref.~\cite{Anselmino:2006yq} by analyzing PHENIX
data for transverse single spin asymmetries for the process
$p^{\uparrow}\,p\to \pi^0\,X$, with the neutral pion being produced in the
central rapidity region.

Clearly, the measurement of $A_N^{\sin\phi_{S}}$ at large $p_{{\rm j} T}$,
where the role of the gluon Sivers function becomes negligible,
could be quite helpful in discriminating between the SIDIS~1 and SIDIS~2
parameterizations and constraining the large $x$ behaviour of the $u$, $d$ quark
Sivers functions.

The present analysis can be extended to the transverse single
spin asymmetry  $A_N^{\sin\phi_S}$ for inclusive jet production in
$p^\uparrow \,p\to {\rm jet}\,X$, by simply integrating the results for
the process $p^\uparrow \,p\to {\rm jet}\,\pi\,X$ over the pion phase space.
In this case, in the general structure of the asymmetry in
Eq.~(\ref{num-asy-gen}), only the
$\sin\phi_S$ modulation will be present, since all the mechanisms related to
the fragmentation process cannot play a role. The numerator of
 $A_N^{\sin\phi_S}$ will be given by Eq.~(\ref{sivers}),
in which the fragmentation function  $D_{1}^c(z,\bm{k}^2_{\perp \pi})$
is replaced  by $\delta(z-1)\,\delta^2(\bm{k}_{\perp \pi})$. As already done
for jet-pion production, we have checked explicitly that,
for the kinematic configurations under study, all other possible contributions
rather than the Sivers one are numerically irrelevant and therefore can be
safely neglected.

\begin{figure*}[t]
 \includegraphics[angle=0,width=0.4\textwidth]{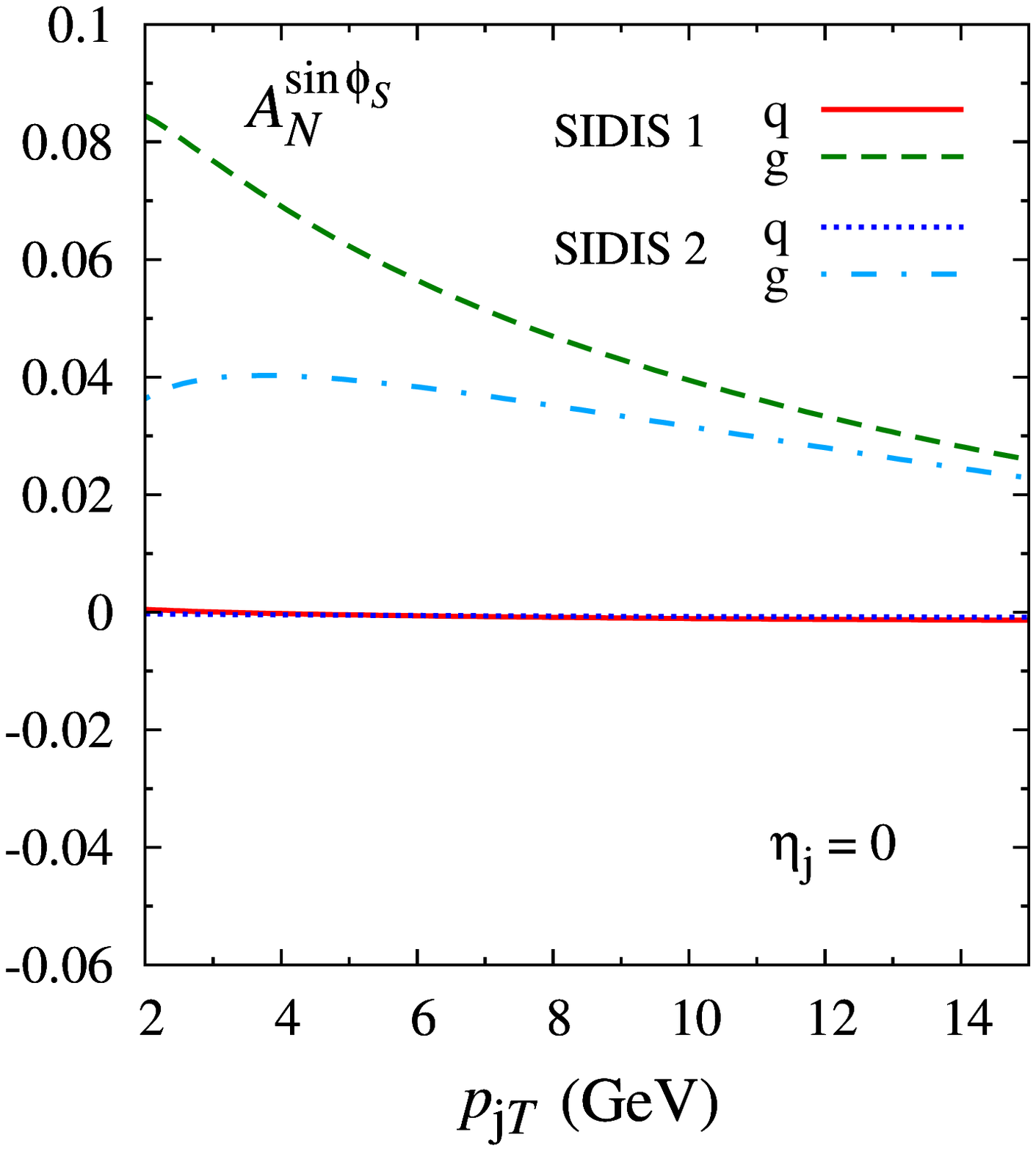}
 \includegraphics[angle=0,width=0.4\textwidth]{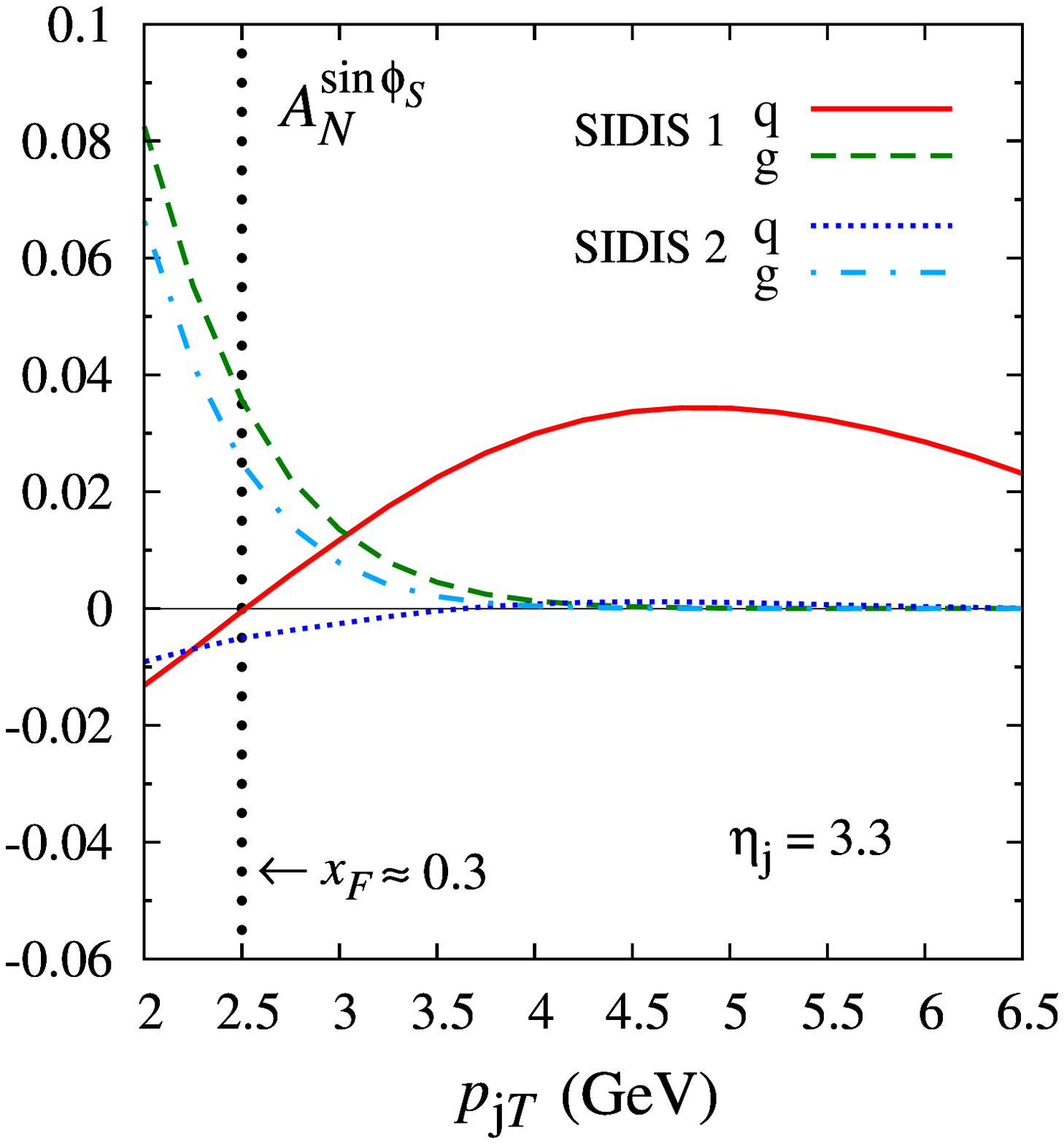}
 \caption{The Sivers asymmetry  $A_N^{\sin\phi_{S}}$
 for the process $p^\uparrow \, p\to {\rm jet}\, X$, as a function
of $p_{{\rm j} T}$, at fixed value of the rapidity $\eta_{\rm j}$ and
c.m.\ energy  $\sqrt{s}= 200$ GeV. Estimates for the quark contribution
are obtained
by adopting the parametrization sets SIDIS~1 and SIDIS~2. The gluon Sivers
function is assumed to be positive and to saturate an updated version of the
bound in Ref.~\cite{Anselmino:2006yq}. The dotted vertical line delimits
the region $x_F\approx 0.3$, beyond which  the currently available
parameterizations for the
quark Sivers function, extracted from SIDIS data, are affected by large
uncertainties.
\label{asy-siv-jet-200} }
\end{figure*}

In Fig.~\ref{asy-siv-jet-200} we present our
results for $A_N^{\sin\phi_S}$ for inclusive jet production
at the c.m.\ energy $\sqrt{s}=200$ GeV,
as a function of $p_{{\rm j} T}$ and fixed rapidities $\eta_{\rm j}=0$
(left panel) and $\eta_{\rm j} = 3.3$ (right panel).
As before, they have been obtained utilizing the parameterizations SIDIS~1
and SIDIS~2 for the quark Sivers functions and an updated version of the
bound presented in Ref.~\cite{Anselmino:2006yq}  for the gluon Sivers
function (taken to be positive). Predictions in the forward rapidity region
are very similar to those for jet-neutral pion production shown in the
central panel of Fig.~\ref{asy-an-siv-par200},
where the gluon component dominates only at very low values of $p_{{\rm j} T}$
and decreases quickly as $p_{{\rm j} T}$ increases. On the other hand,
in the central rapidity region, the gluon component is always larger than
the quark one, the latter being practically negligible.
A measurement of $A_N^{\sin\phi_S}$  in this kinematic
domain would therefore be ideal to probe the gluon Sivers function \cite{D'Alesio:2010am,Adamczyk:2012qj}.
Results for RHIC kinematics at $\sqrt{s}=500$ GeV will be discussed in the next section.

Notice that the scan procedure discussed in section~\ref{sec:res-coll} and in Ref.~\cite{Anselmino:2012rq}
for the Collins effect in the process $p^\uparrow p\to h\,X$, and for the large $x$ behaviour of
the transversity distribution, can also be applied, for the same process, to the Sivers
asymmetry and, in this case, the large $x$ behaviour of the Sivers distribution, see Ref.~\cite{Anselmino:2013rya}.
We will present some new results obtained utilizing the scan procedure for the Sivers azimuthal
asymmetry $A_N^{\sin\phi_S}$ in $p^\uparrow p\to {\rm jet}\,\pi\,X$ and  $p^\uparrow p\to {\rm jet}\,X$ processes
in the next section.

\section{A study of the process dependence of the Sivers function}

In the GPM approach adopted so far TMD distribution and fragmentation
functions are assumed to be universal.
 In particular, the Sivers function
in Eq.~(\ref{sivers}) is taken to be the same as the one extracted from SIDIS~\cite{Anselmino:1994tv, D'Alesio:2010am},
\begin{equation}
f_{1T}^{\perp a}(x_a, \bm{k}_{\perp a}^2) \equiv f_{1T}^{\perp a, \rm SIDIS}(x_a, \bm{k}_{\perp a}^2).
\end{equation}

There is at present a large consensus on the universality of the Collins
fragmentation function (which however must be verified phenomenologically),
at least for processes where QCD factorization has been proven.
On the contrary, several naively time-reversal odd (T-odd) TMD distributions
crucially depend on initial and/or final state interactions (embedded via gauge links)
among struck partons and soft remnants in the process.

Recently the azimuthal asymmetries for the distribution of leading pions
inside jets have been studied allowing for the process dependence of the quark
Sivers function \cite{D'Alesio:2011mc} within the framework of the so-called
colour gauge invariant GPM  \cite{Gamberg:2010tj}. In the CGI~GPM the
existence of a nonzero Sivers function in a transversely polarized hadron is
due to the effects of initial (ISIs) and final (FSIs) state interactions
between the struck parton and the spectator remnants from the polarized proton.
These interactions  depend on the particular process considered and make
the Sivers function non-universal. The typical example is provided by the
predicted opposite sign of the quark Sivers functions in SIDIS, where only
FSIs are present, and in the DY process, in which only ISIs can be active.

\begin{figure}[t]
\includegraphics[angle=0,width=0.35\textwidth]{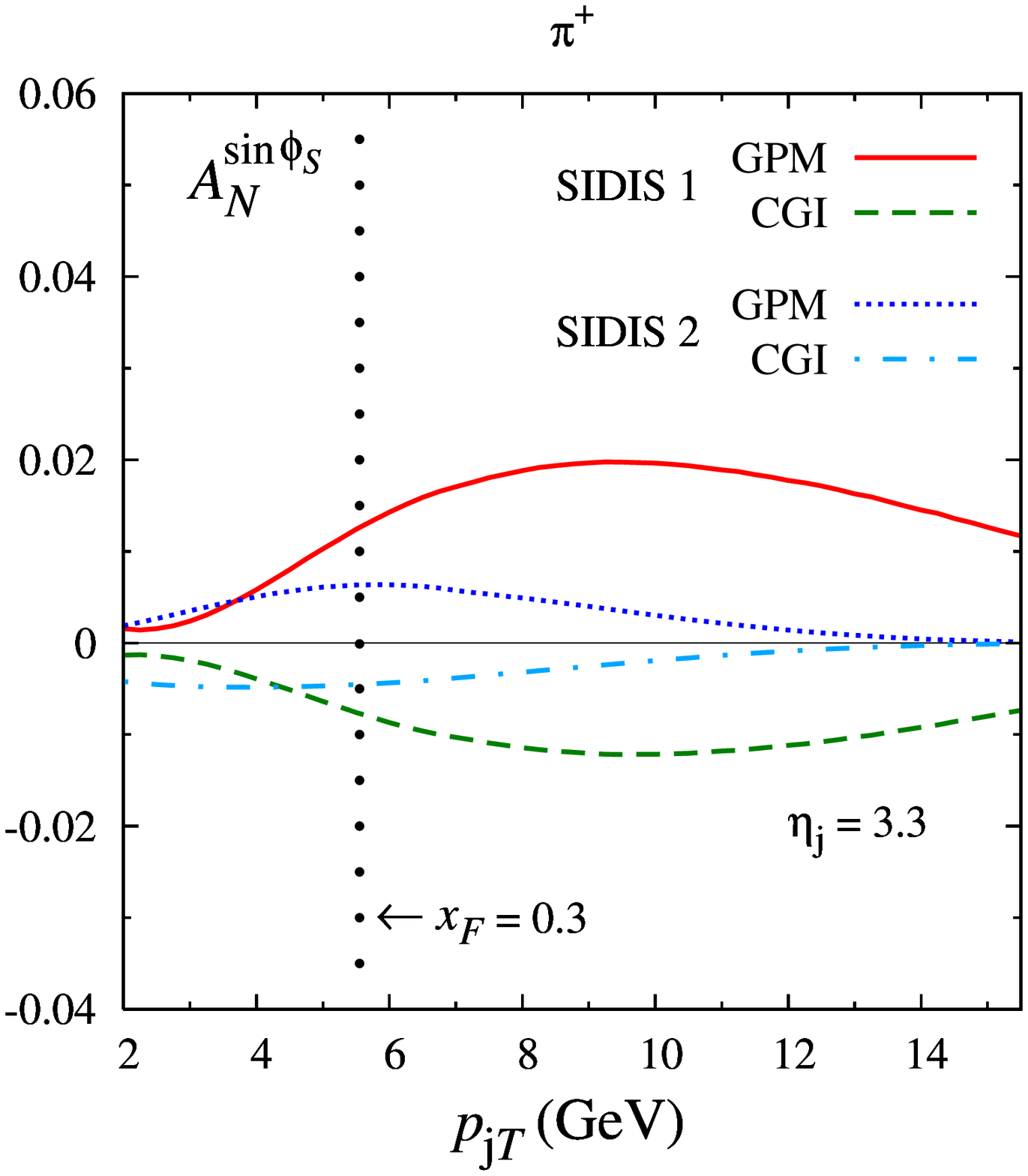}
 \hspace*{-20pt}
 \includegraphics[angle=0,width=0.35\textwidth]{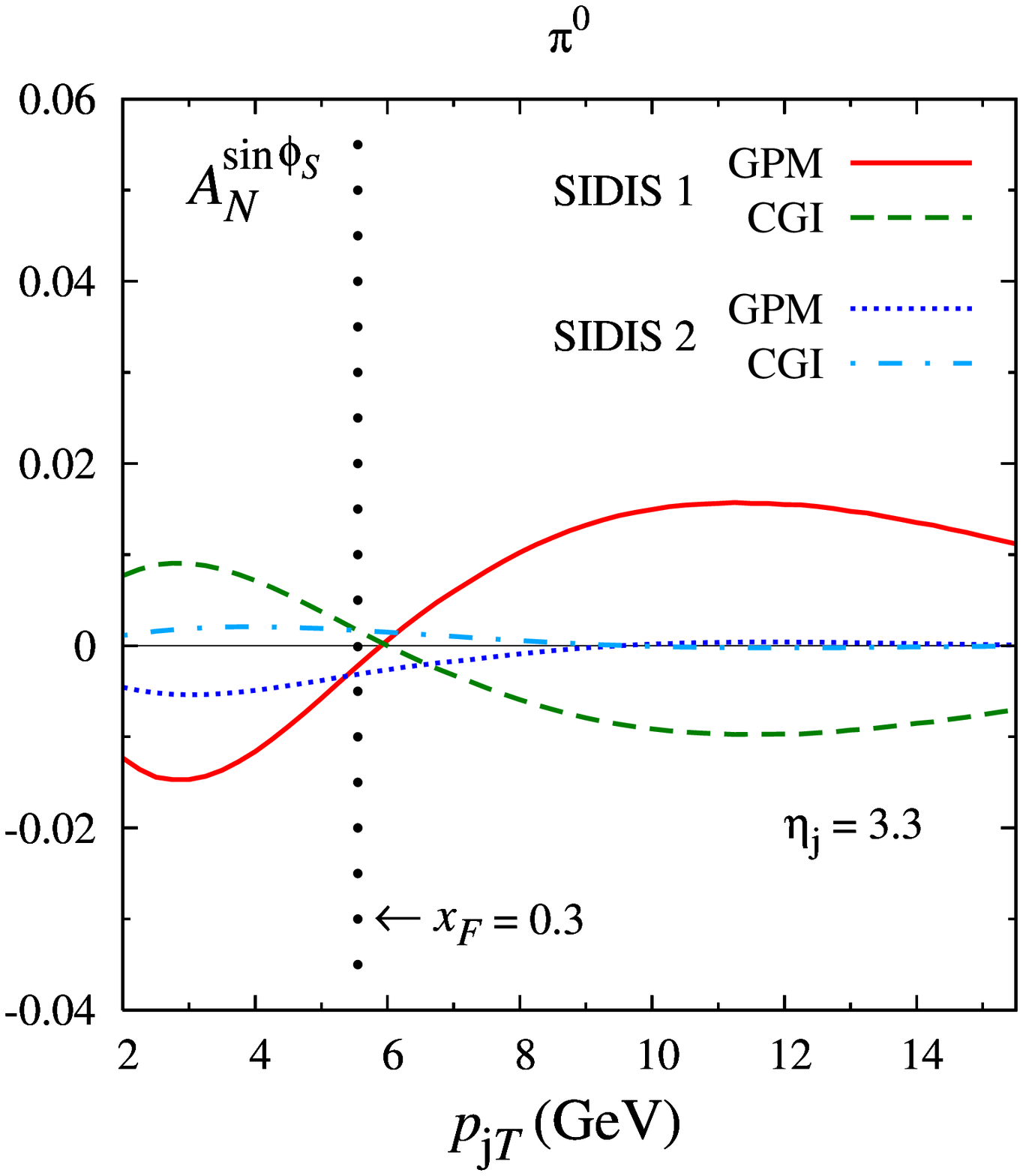}
 \hspace*{-20pt}
 \includegraphics[angle=0,width=0.35\textwidth]{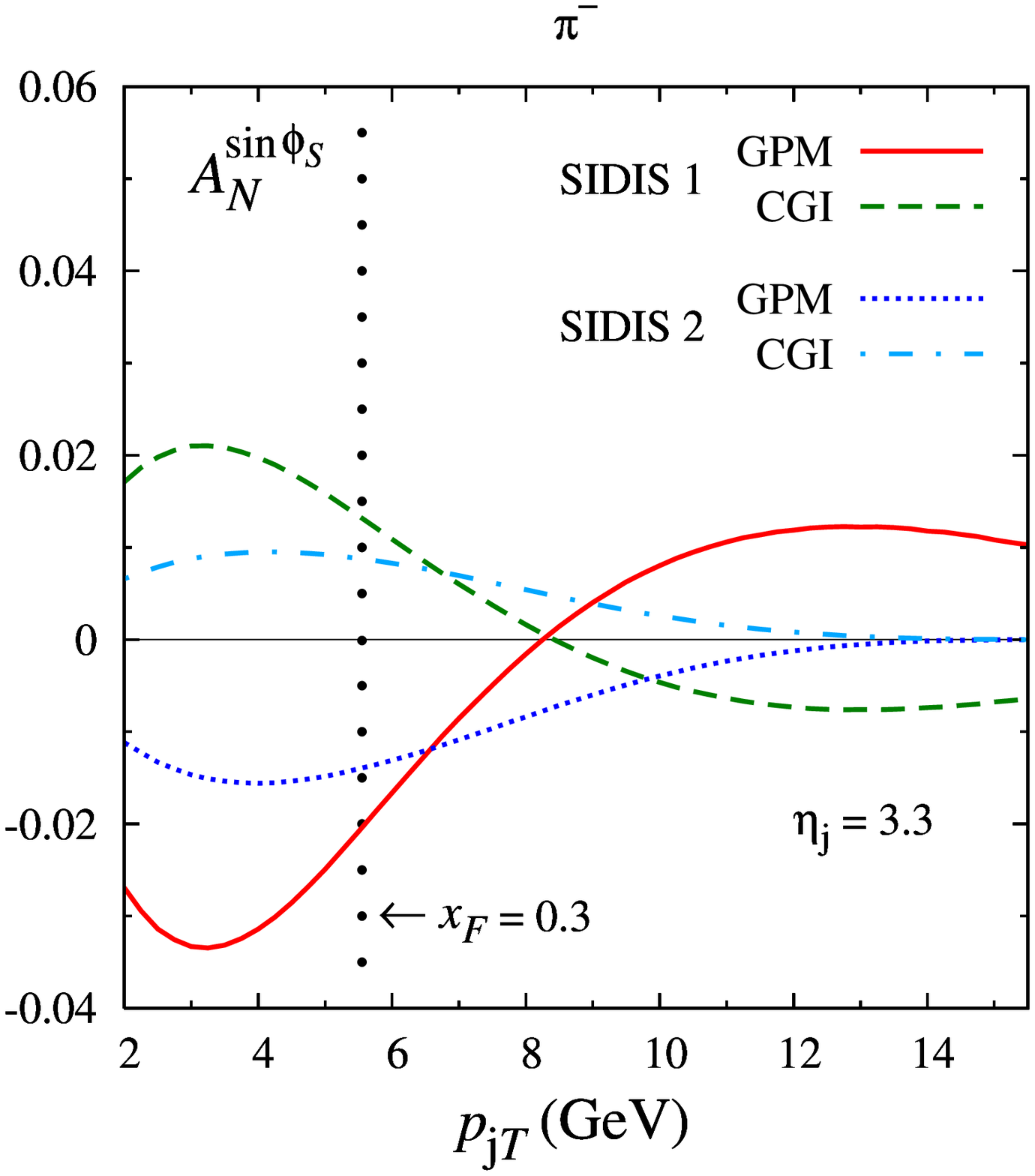}
\caption{The quark contribution to the Sivers asymmetry
$A_N^{\sin\phi_{S}}$ in the GPM and CGI~GPM approaches  for the process $p^\uparrow \, p\to {\rm jet}\, \pi\,X$,
as a function of $p_{{\rm j} T}$, at fixed value of the rapidity $\eta_{\rm j}$ and  c.m.\ energy  $\sqrt{s}= 500$ GeV.
Estimates are obtained
by adopting the parametrization sets SIDIS~1 and SIDIS~2. The dotted
vertical line delimits the region $x_F\approx 0.3$, beyond which  the
currently available parameterizations for the quark Sivers function,
extracted from SIDIS data, are affected by large uncertainties.}
\label{fig1}
\end{figure}

The colour factor structure of the Sivers function for the reaction under study,
involving hadrons in both the initial and the final states,
is more complicated because both ISIs and FSIs contribute. Eq.~(\ref{sivers}) has then to be replaced by
\begin{eqnarray}
{E_{\rm j}}\, \frac{\de\Delta\sigma^{(\rm Sivers)}}{\de^3 {\bm p}_{\rm j}\, \de z\, \de^2 {\bm k}_{\perp \pi}}
& = & \frac{2\, \alpha_s^2}{s} \sum_{a,b,c,d} \int \frac{\de x_a}{x_a}\,\de^2\bm{k}_{\perp a}\,
\int \frac{\de x_b}{x_b}\,\de^2{\bm k}_{\perp b} \,\delta(\hat s+\hat t+\hat u)\,H^{U}_{ab\to cd}(\hat s,\hat t,\hat u)
\nonumber \\
&&\qquad\times
\Big ( -\frac{k_{\perp a}}{M}\Big)f_{1T}^{\perp a, ab\to cd}(x_a, {\bm k}_{\perp a}^2) \cos\phi_a
\, f_{b/B}(x_b, {\bm k}_{\perp b}^2)\, D_{1}^c(z, {\bm k}_{\perp \pi}^2) \sin\phi_{S} \,,
\label{process}
\end{eqnarray}
in which a {\it process-dependent} Sivers function denoted as $f_{1T}^{\perp a, ab\to cd}$ is used.
The resulting colour factors, $C_I$ ($C_{F_c}$), for initial (final) state interactions determine the proper Sivers function to be used for each of
the different partonic scattering processes $a\, b\to c\, d$.
They are the same as
the  ones calculated in Ref.~\cite{Gamberg:2010tj} for single inclusive hadron production using a one-gluon exchange approximation. Finally, the process dependence of the Sivers function can be absorbed into the squared hard partonic
scattering amplitude $H^{U}_{ab\to cd}$, that is
\begin{equation}
f_{1T}^{\perp a, ab\to cd} \,H^{U}_{ab\to cd} \equiv f_{1T}^{\perp a, \rm SIDIS} \,H^{\rm Inc}_{ab\to cd}\,,
\label{HInc}
\end{equation}
where the new hard function $H^{\rm Inc}_{ab\to cd}$ has been introduced.
Details on the connection between the CGI~GPM and the twist-three
collinear  formalism~\cite{Qiu:1991pp, Kouvaris:2006zy}, suggested by Eq.~(\ref{HInc}),
can be found in Ref.~\cite{Gamberg:2010tj}.
\begin{figure}[t]
\includegraphics[angle=0,width=0.35\textwidth]{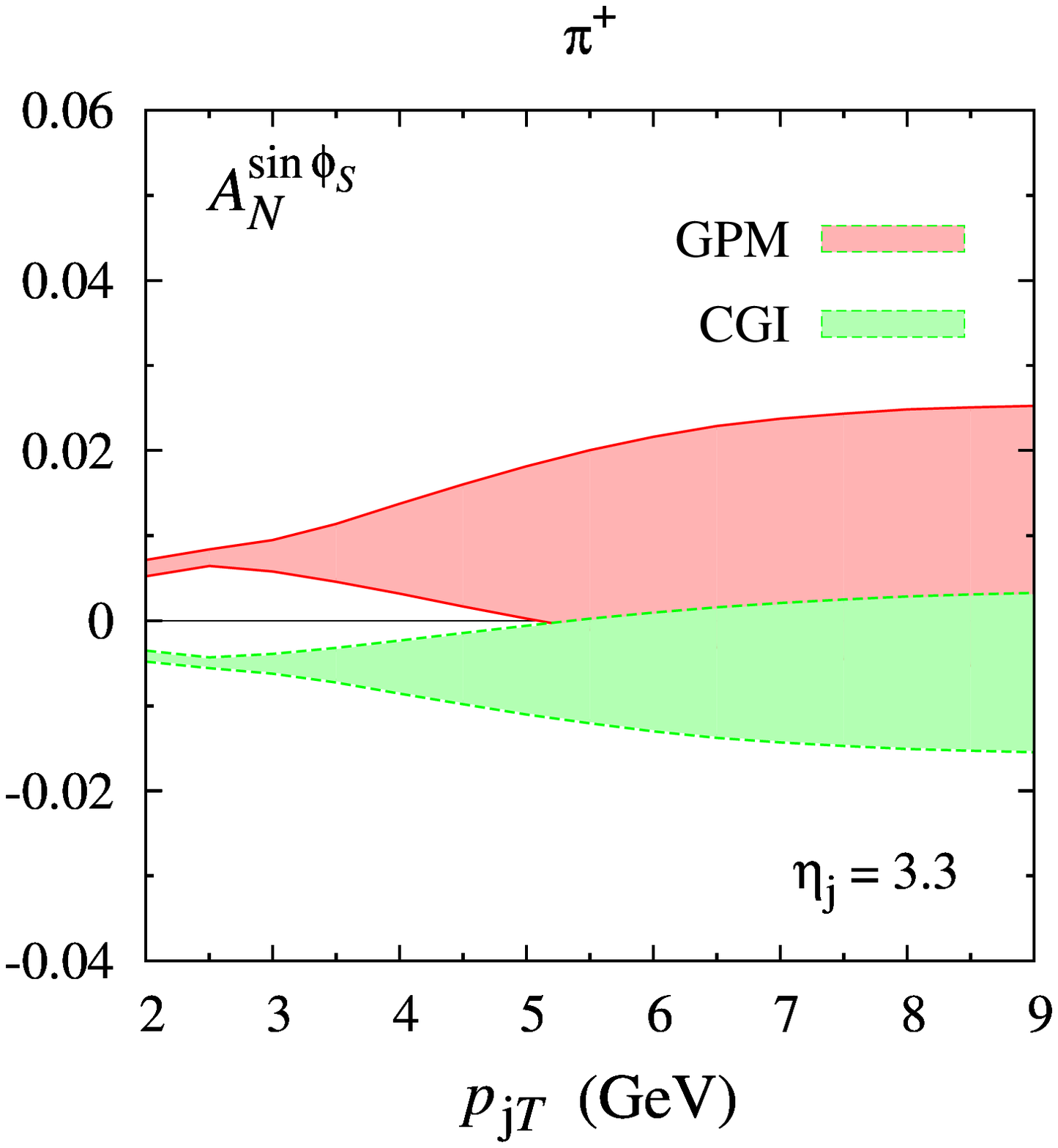}
 \hspace*{-20pt}
 \includegraphics[angle=0,width=0.35\textwidth]{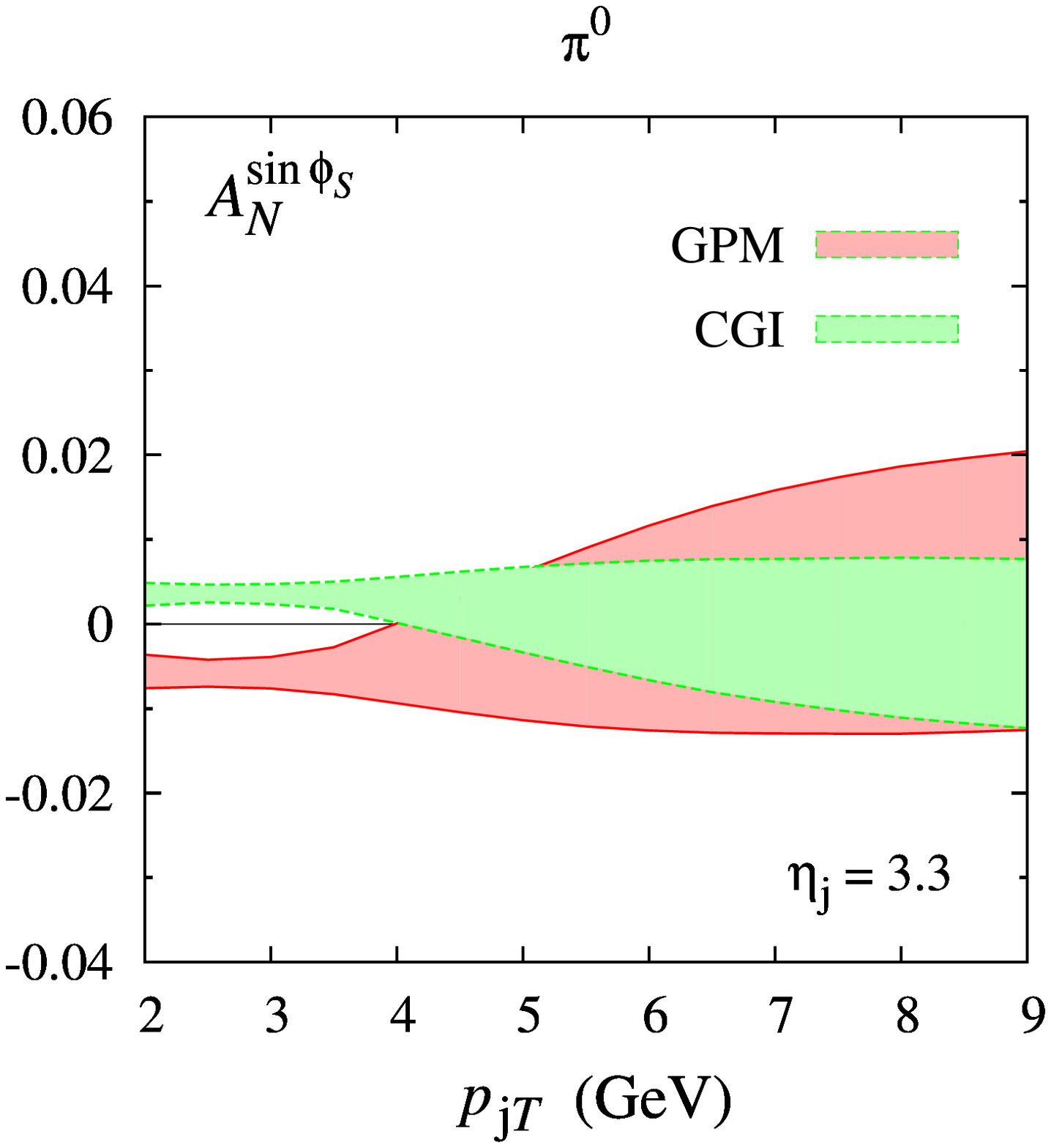}
 \hspace*{-20pt}
 \includegraphics[angle=0,width=0.35\textwidth]{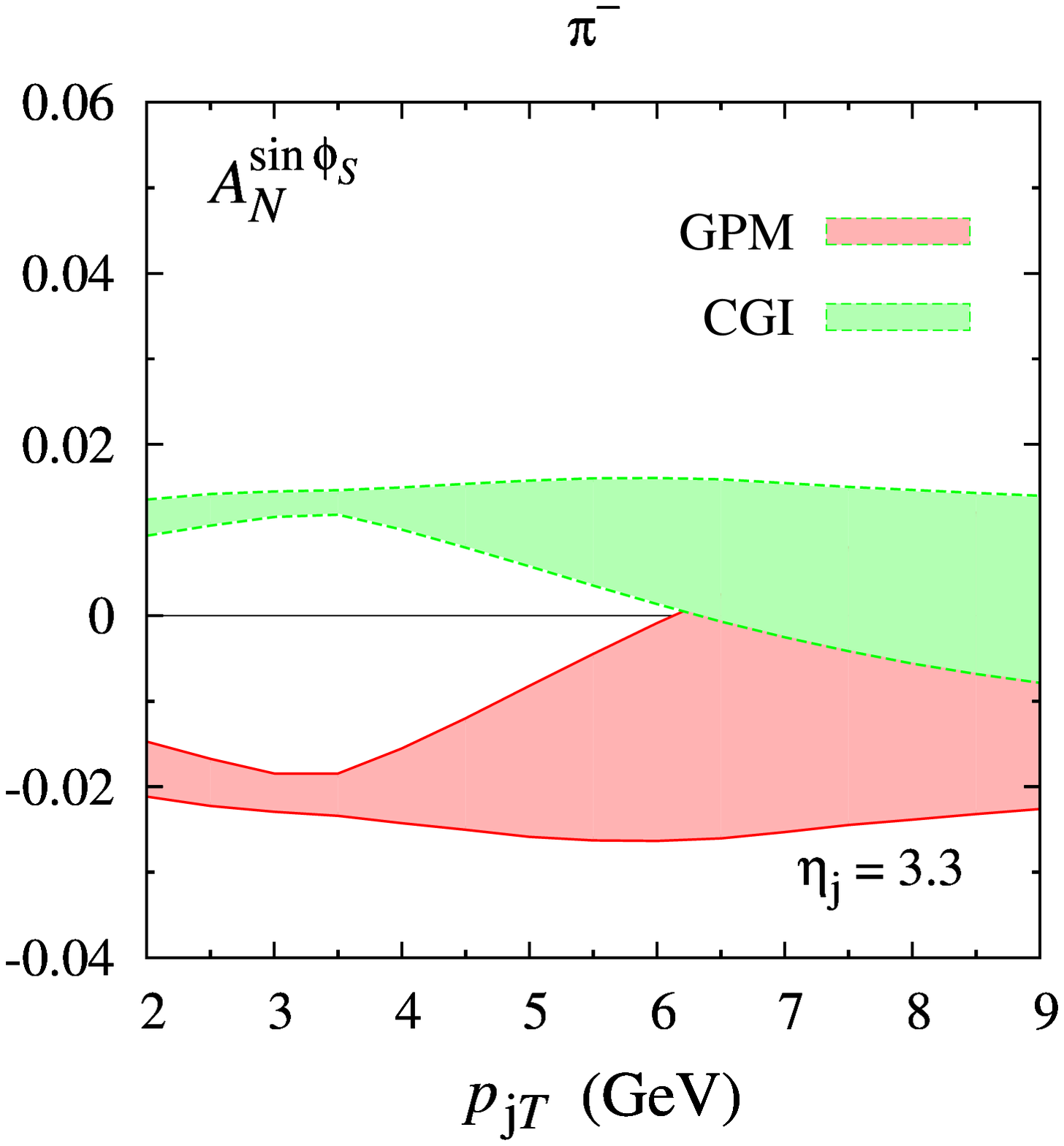}
\caption{
Scan bands (that is, the envelope of possible values) for the
quark contribution to the Sivers asymmetry
$A_N^{\sin\phi_{S}}$  in the GPM and CGI~GPM approaches, for the process $p^\uparrow \, p\to {\rm jet}\, \pi\,X$,
as a function of $p_{{\rm j} T}$, at fixed value of the rapidity $\eta_{\rm j}$ and  c.m.~energy  $\sqrt{s}= 500$ GeV.
  The shaded bands are generated following the scan procedure explained in the text
  (see Refs.~\cite{Anselmino:2012rq,Anselmino:2013rya} for more details).}
\label{asy-siv-scan-500}
\end{figure}

Since our aim is to study the process dependence of the quark Sivers function,
we analyze  pion-jet production in the forward rapidity region, where possible
contributions from sea-quark and gluon Sivers functions are expected to be
negligible.   This assumption is supported by studies of SSAs in SIDIS~\cite{Anselmino:2008sga} and in $pp\to \pi\, X$
 processes at central rapidities~\cite{Anselmino:2006yq, Adler:2005in, Wei:2011nt} and by the analysis performed in Ref.~\cite{Brodsky:2006ha}. Our results are shown in Fig.~\ref{fig1},
where $A_N^{\sin\phi_{S}}$, integrated over $\bm{k}_{\perp\pi}$ and $z$ ($z\ge 0.3$), is plotted as a function of the jet transverse momentum $p_{{\rm j} T}$ at fixed jet rapidity $\eta_{\rm j}=3.3$, for  the RHIC energy $\sqrt{s}=500$ GeV.
The solid and dotted lines represent our predictions in the GPM formalism
using the two available sets, SIDIS~1 and SIDIS~2 respectively, for the quark Sivers function, while the dashed and dot-dashed lines describe the
analogous predictions in the CGI~GPM formalism. As one can easily see, the results obtained with and without inclusion of colour gauge factors are comparable in size but have {\em opposite signs} \cite{D'Alesio:2011mc}, in close analogy to the DY case. The reason is that, at forward rapidity, the dominant channel is
$qg\to qg$, where the final quark is identified with the observed jet, for which
 the effects of  ISIs/FSIs lead to
\begin{equation}
H^{\rm Inc}_{qg\to qg}\sim -\frac {N_c^2+2}{N_c^2-1}\,\frac{\hat s^2}{\hat t^2}
\end{equation}
in the CGI~GPM, while
\begin{equation}
H^{U}_{qg\to qg}\sim \frac{2\hat s^2} {\hat t^2}
\end{equation}
in the GPM. Moreover, as already pointed out in the previous section, our
estimates obtained adopting the two different parameterizations SIDIS~1 and
SIDIS~2 are similar only in the region $x_F \le 0.3$, corresponding to
 $p_{{\rm j} T}\le 5.5$ GeV at $\sqrt{s}=500$ GeV. Therefore this is the
optimal  kinematic region to test directly the process dependence of the
Sivers function:~the measurement of a sizable asymmetry for
$p_{{\rm j} T}\le 5.5$ GeV could easily  discriminate between the two
different approaches and probe the universality properties of the Sivers
function. At the c.m.\ energy  $\sqrt s = 200$ GeV  our predictions would be
qualitatively similar to the ones presented in Fig.~\ref{fig1},
becoming almost twice as large. However the range of $p_{{\rm j} T}$
covered would now be narrower, $p_{{\rm j} T}\le 6.5$~GeV, and $x_F \le 0.3$
would correspond to $p_{{\rm j} T}\le 2.5$~GeV.

As already discussed in the previous section for the Collins azimuthal asymmetry,
the scan procedure introduced in Refs.~\cite{Anselmino:2012rq,Anselmino:2013rya}
offers a different and more complete information. In fact, it gives the envelope of all
possible values of $A_N^{\sin\phi_S}$ coming
from parameterizations of the Sivers function leading to good fits of the SIDIS
data on the analogous asymmetry.
Therefore, in Fig.~\ref{asy-siv-scan-500} we present the analogous of Fig.~\ref{fig1}
obtained using new results of the Sivers scan procedure.
These plots confirm the conclusions drawn from Fig.~\ref{fig1}: the low-intermediate $p_{{\rm j} T}$
region is the most interesting for a discrimination between the GPM and CGI~GPM approaches.
As soon as $p_{{\rm j} T}$ grows beyond $4-6$ GeV the two scan bands start overlapping
and we loose predictive power. For this reason, we cut our plots at $p_{{\rm j} T}=9$ GeV, although
the kinematical limit is larger (see Fig.~\ref{fig1}).
Clearly, the most favourable situation seems to be that of the $\pi^-$,
for which the asymmetry is larger and the scan bands for the GPM and CGI~GPM cases are
well separated up to $p_{{\rm j} T}\simeq 5$ GeV.

\begin{figure}[t]
\begin{center}
 \includegraphics[angle=0,width=0.4\textwidth]{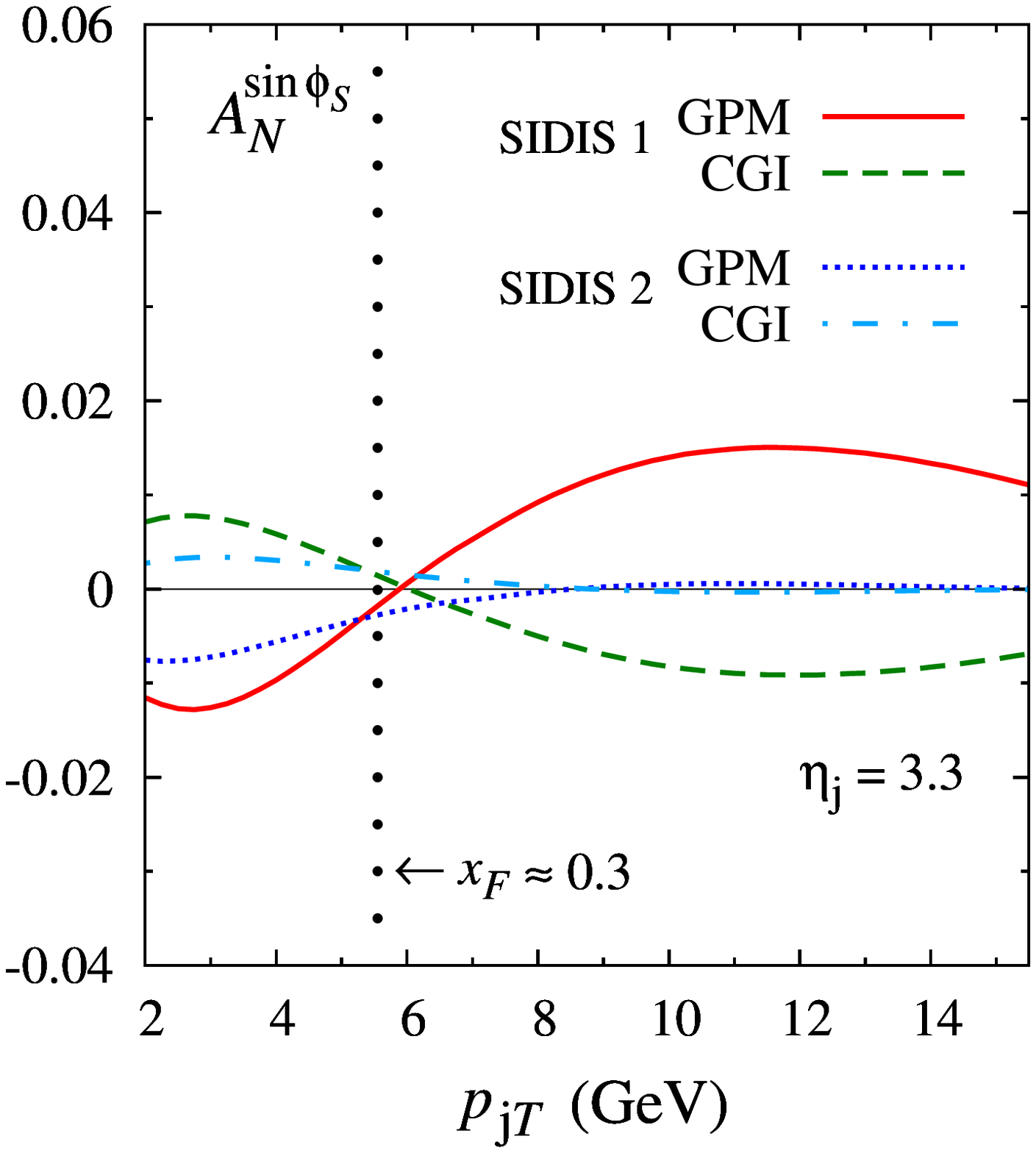}
 \hspace*{0pt}
 \includegraphics[angle=0,width=0.4\textwidth]{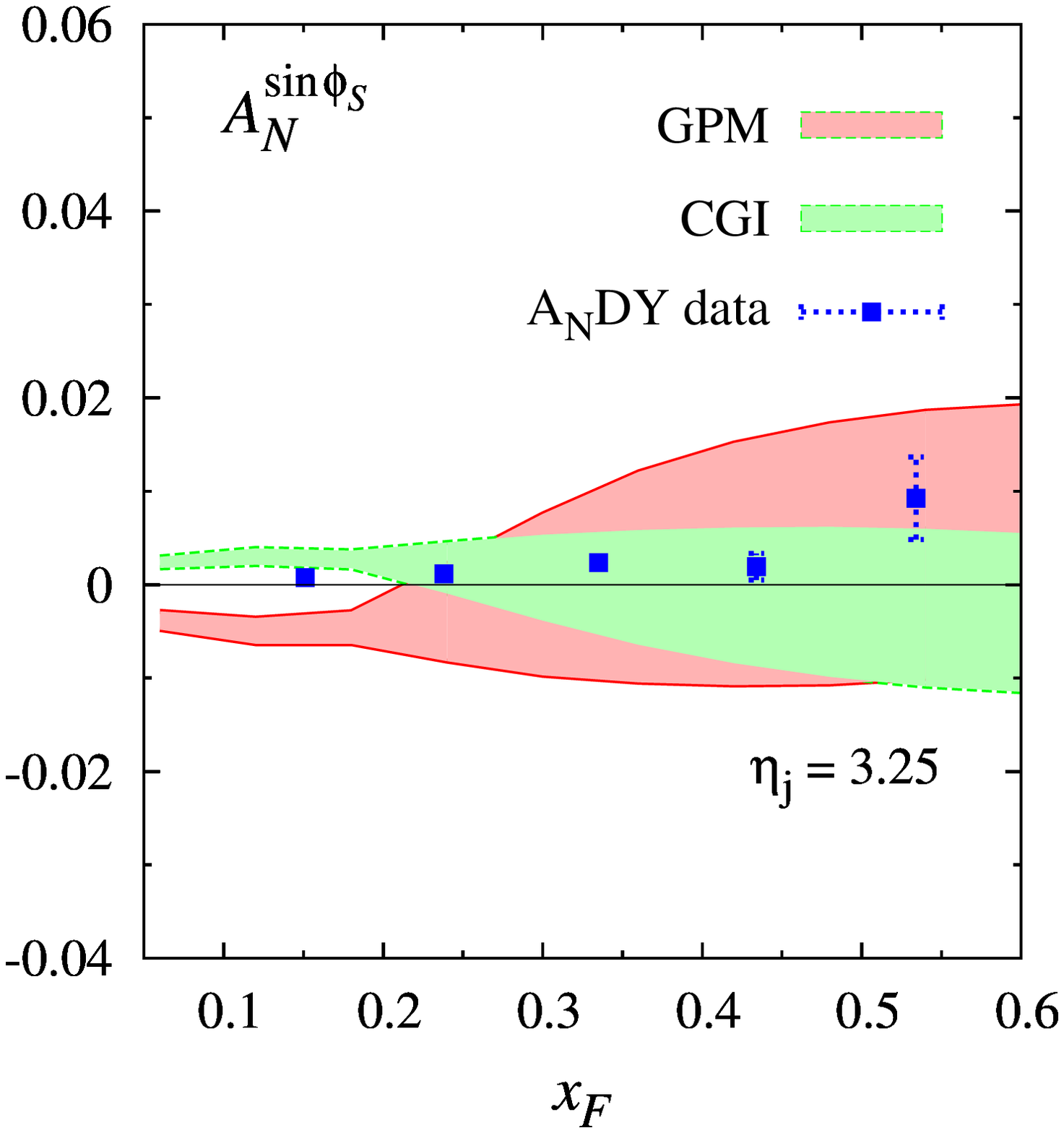}
 \caption{
 Left panel: The quark contribution to the Sivers asymmetry
$A_N^{\sin\phi_{S}}$ in the GPM and CGI~GPM approaches  for the process $p^\uparrow \, p\to {\rm jet}\,X$,
as a function of $p_{{\rm j} T}$, at fixed value of the rapidity $\eta_{\rm j}=3.3$ and  c.m.\ energy  $\sqrt{s}= 500$ GeV.
Estimates are obtained by adopting the parametrization sets SIDIS~1 and SIDIS~2. The dotted
vertical line delimits the region $x_F\approx 0.3$, beyond which  the
currently available parameterizations for the quark Sivers function,
extracted from SIDIS data, are affected by large uncertainties.
Right panel: Scan bands (that is, the envelope of possible values) for the
quark contribution to the Sivers asymmetry
$A_N^{\sin\phi_{S}}$  in the GPM and CGI~GPM approaches, for the process $p^\uparrow \, p\to {\rm jet}\,X$,
as a function of $x_F$, at fixed value of the rapidity $\eta_{\rm j}=3.25$ and  c.m.~energy  $\sqrt{s}= 500$ GeV.
  The shaded bands are generated following the scan procedure explained in the text
  (see Refs.~\cite{Anselmino:2012rq,Anselmino:2013rya} for more details).}
 \label{asy-an-siv-jet-par500}
\end{center}
\end{figure}

Finally, we consider also single inclusive jet production in proton-proton
 scattering. Data for this observable are now available and
 have been presented in  Refs.~\cite{Bland:2013pkt,Nogach:2012sh}.
  The results obtained for $A_N^{\sin\phi_{S}}$  are plotted in
Fig.~\ref{asy-an-siv-jet-par500}. In the left panel, we show $A_N^{\sin\phi_{S}}$
for the $p^\uparrow p\to {\rm jet}\,X$ process, as a function of $p_{{\rm j} T}$ and
fixed pseudorapidity, $\eta_{\rm j}=3.3$, at RHIC c.m.~energy $\sqrt{s}=500$ GeV.
The results look very similar to those for the
case of neutral pion-jet production, shown in the central panel of
Fig.~\ref{fig1}.
In the right panel of Fig.~\ref{asy-an-siv-jet-par500} we compare
the GPM and CGI~GPM scan bands for the Sivers asymmetry $A_N^{\sin\phi_{S}}$
with recent results by the A$_N$DY Collaboration~\cite{Bland:2013pkt,Nogach:2012sh},
shown as a function of $x_F$, at fixed pseudorapidity
$\eta_{\rm j}=3.25$ and $\sqrt{s}=500$ GeV.
As expected, beyond $x_F\sim 0.3$, since the $u$, $d$ quark Sivers functions
are poorly constrained by present SIDIS data, the scan bands become larger and overlap
almost completely. Therefore, at this stage we cannot draw any conclusion by looking solely
at these results. Only the first and the last A$_N$DY data points seems to favour
respectively the CGI~GPM and the GPM approach, but much more work is needed.
See also Ref.~\cite{Gamberg:2013kla} for a similar study comparing the GPM
and collinear twist-three results.

\section{\label{sec-other-tests}
Other tests of the process dependence of the TMD functions}

In this section we present a short overview of other possible tests of
the process dependence of TMD
parton distribution and fragmentation functions proposed in the literature.
Due to lack of space, we will not cover thoroughly all aspects of the subject, limiting
ourselves to a discussion of the more interesting phenomenological tests.
A detailed treatment may be found in the original papers quoted in the bibliography.

Basically, all these phenomenological studies try to compare predictions for
spin asymmetries coming from different formalisms (like the collinear twist-three,
the GPM and CGI GPM approaches) in kinematical situations where
typically only one of the many possible effects dominates. If the predictions
of the various approaches are very different (in particular, in sign) then
interesting phenomenological investigations can be performed.

In Ref.~\cite{Bacchetta:2007sz} it was proposed to study a
weighted asymmetry in the azimuthal distribution of photon-jet pairs in the
polarized process $p^\uparrow p\to \gamma\, {\rm jet}\, X$.
It was shown that for specific kinematical configurations reachable
at RHIC, the asymmetry is dominated by the quark Sivers effect, making
its interpretation much more clear. Moreover, predictions
coming from gluonic-pole cross sections~\cite{Bacchetta:2005rm},
directly related to the Wilson lines preserving colour gauge invariance
and leading to process dependent effects,
are almost opposite to those of the generalized parton model.
Therefore, experimental tests of these results offer an interesting alternative
way to investigate the process dependence of the Sivers function and the
predicted relative sign difference in SIDIS and Drell-Yan processes.

As we have already discussed in the previous section,
in Ref.~\cite{Gamberg:2010tj} Gamberg and Kang
have discussed a modified version of the generalized parton model, the
colour gauge invariant GPM. Assuming, as in the GPM, the validity of factorization
for single inclusive particle production in hadronic collisions, this approach
includes the process dependence of TMDs by taking into account initial and
final state interactions between the struck parton and the parent hadron remnants.
Once more, these interactions come out from appropriate, process-dependent colour gauge links.
It was also shown that the CGI GPM is in close connection with the collinear twist-three
approach. The phenomenological implications of the CGI GPM for the process dependence
of the Sivers effect in $p^\uparrow p\to\pi^0, \gamma + X$ reactions were investigated.
Once again, the main result is that the transverse single spin asymmetry due to
the quark Sivers contribution has a similar size but opposite sign with respect to the
original GPM that assumes the universality of TMDs.
Applications of the approach to pion-jet production were discussed in the previous section and in more
detail in Ref.~\cite{D'Alesio:2011mc}.

The study of the universality and process dependence of the Sivers function is
of relevance also in the context of the so-called ``sign mismatch" issue
for the collinear twist-three approach~\cite{Kang:2011hk}.
Since in this formalism factorization has been proven for both
SIDIS processes and single inclusive particle production
in hadronic collisions at large energy scales, the multi-parton soft correlation
functions involved are universal and process independent.
On the other hand, factorization holds also for the TMD approach
in SIDIS, for large $Q^2$ and small transverse momentum of the final hadron.
It has been shown that there is a common region of validity of these two
approaches, and this allows to find a relation among the twist-three
quark-gluon correlation function and the first $\bm{k}_\perp$ moment
of the TMD Sivers function.
However, if one uses this relation from SIDIS processes for the
calculation in the twist-three approach of $A_N$ in $p^\uparrow p\to\pi^0, \gamma + X$
processes, one finds results opposite in sign with respect to those obtained
by directly fitting, in the same approach, the RHIC data for  $p^\uparrow p\to\pi^0\, X$.

In Ref.~\cite{Kang:2012xf} the authors have explored the possibility
of escaping this sign-mismatch problem for the twist-three approach
by accounting for nodes of the quark Sivers function
(either in its $x$ or $\bm{k}_\perp$ dependence).
They found that by allowing for a single node in the quark Sivers function
one is not able to ``cure" the sign mismatch problem and explain both the STAR
and BRAHMS $A_N$ data for $p^\uparrow p\to \pi\,X$ reactions.
However, one must not forget that the Sivers effect is not the
only possible contribution to $A_N$.
In fact, it may be that the Sivers effect gives a subdominant
contribution, and the asymmetry is mainly due to the Collins effect
in the fragmentation sector.
To investigate this eventuality it is crucial to collect experimental information
for processes like, e.g., $p^\uparrow p\to\gamma\, X$ and  $p^\uparrow p\to {\rm jet}\, X$
where fragmentation in the final state is absent.

As we have seen, quite recently the A$_N$DY Collaboration at
RHIC~\cite{Bland:2013pkt,Nogach:2012sh} has presented preliminary results for
$A_N(p^\uparrow p\to {\rm jet}\, X)$ at  forward rapidity and c.m.~energy $\sqrt{s}=500$ GeV.
Gamberg, Kang and Prokudin~\cite{Gamberg:2013kla} have performed a new fit of the Sivers
function using HERMES and COMPASS data on the $A_N^{\sin(\phi_h-\phi_S)}$ asymmetry.
Then, using this information and the relation among the twist-three quark-gluon
correlation function and the first $\bm{k}_\perp$ moment of the Sivers function
discussed above, they have estimated the spin asymmetry for $p^\uparrow p\to {\rm jet}\, X$
in the collinear twist-three approach, comparing it with A$_N$DY data.
They found, taking into account that the large $x$ behaviour of the Sivers
function is poorly constrained by present SIDIS data, that their estimate is consistent
with experimental data and there is in fact no strong sign mismatch problem,
contrary to the case of pion single spin asymmetries discussed above.

Kang and Qiu~\cite{Kang:2009bp} have proposed to probe
the (modified) universality of the quark and gluon Sivers functions,
that is the change of sign between the Sivers functions in SIDIS and DY processes,
by studying the transverse single spin asymmetry $A_N$ for $W$ production and
inclusive lepton production from $W$ decays in polarized proton-proton collisions at RHIC energies.
Although the lepton asymmetry is diluted by the $W$ decays, its size can reach several percents
over a large range of lepton rapidity at RHIC.
Therefore this process can offer an additional phenomenological test of the predicted
sign change of the Sivers function. Moreover, because of the weak interaction,
it can provide unique information, with respect to the DY case, on the flavour
dependence and the functional form of the Sivers function.

Let us finally add some comments on the process dependence of the T-odd TMD fragmentation
functions, like the Collins function and the so-called ``polarizing" fragmentation
function~\cite{Mulders:1995dh,Boer:1997nt,Anselmino:2000vs}.
They have been shown to be universal by several authors,
see e.g.~Refs.~\cite{Collins:2004nx,Meissner:2008yf,Yuan:2007nd,Gamberg:2010uw,Yuan:2009dw}.
Testing phenomenologically universality in the fragmentation sector is as important
as the tests for the modified universality of the Sivers functions discussed above.
However, for the Collins function, the study of its universality is made difficult by its
chiral-odd nature. In any physical observable it will always appear coupled to
another chiral-odd object, either in the distribution or in the fragmentation
sector. As well known examples, the Collins function couples to the TMD
transversity distribution in SIDIS and in the pion-jet production process considered
in detail here. It couples to another Collins FF in $e^+e^-\to h_1 h_2 \,X$ reactions.
Therefore, relative signs among these coupled chiral-odd functions are
difficult to determine and require the study of different observables involving
additional chiral-odd functions.

Based on these considerations, the authors of Ref.~\cite{Boer:2010ya}
have suggested to study the universality of the polarizing fragmentation
functions and test factorization by looking at the transverse polarization of
$\Lambda$ hyperons in SIDIS processes and $e^+e^-$ annihilations.
They found that, despite the large uncertainties in these functions,
definite signs for the hyperon polarization in different processes
can be obtained, possibly allowing for a robust test of universality
in this sector.

\section{\label{sec-conclusions} Conclusions}

In the last years, impressive progress has been made in the theoretical understanding of the origin
of the sizable azimuthal and spin asymmetries measured by several experiments in polarized
hadronic processes at large energy scales. The crucial role of colour gauge invariance, and of
the proper account of gauge links (Wilson lines) also in the transverse plane with
respect to the usual light-cone direction, has been emphasized and investigated in depth.
Several processes and polarized observables, for which factorization may not hold and
universality can be broken, have been recognized.
However, it is always difficult to assess, for the ongoing experiments, as well as
for the ones which are going to be performed in the near future, the real relevance and size of
process-dependent terms and factorization-breaking effects.
Clearly, theoretical, more formal,  developments must be complemented by corresponding detailed
phenomenological analyses. These can be of great help and valuable guidance for further
theoretical progress in this field.

In this review we have discussed, in the framework of the so-called generalized
parton model, the phenomenological relevance and usefulness of the reaction
$p^\uparrow p\to{\rm jet}\,\pi\,X$ for the study of the process dependence of the
TMD PDFs and FFs, in particular for the Sivers distribution and the Collins fragmentation function.
We have shown how the study of this process can well complement information coming from SIDIS, Drell-Yan and
$e^+e^-$ annihilations, particularly for the knowledge of the large $x$ behaviour of the TMD quark
transversity distributions and of the quark Sivers functions.
We have also shortly summarized additional phenomenological tests, formulated within various
theoretical approaches, recently suggested in the literature for the study of
the universality properties and the process dependence of the TMDs.

\begin{acknowledgments}
We acknowledge financial support from the European Community under the FP7
``Capacities - Research Infrastructures'' programme (HadronPhysics3, Grant Agreement 283286).
U.D.~and F.M.~acknowledge partial support by Italian Ministero dell'Istruzione,
dell'Universit\`{a} e della Ricerca Scientifica (MIUR) under Cofinanziamento PRIN 2008.
U.D. is grateful to the Department of Theoretical Physics II of the Universidad Complutense
of Madrid for the kind hospitality extended to him during the completion of this work.
\end{acknowledgments}


\end{document}